\documentclass[twocolumn]{autart}    
\usepackage{color,latexsym,amsfonts,amssymb}
 \usepackage{amsmath}
  \usepackage{dsfont}
\usepackage{graphicx}          
 \usepackage[round]{natbib}
\newtheorem{theorem}{Theorem}[section]
\newtheorem{lemma}[theorem]{Lemma}
\newtheorem{remark}{Remark}[section]

\newtheorem{corollary}[theorem]{Corollary}

\newtheorem{assumption}{Assumption}[section]
\newtheorem{definition}{Definition}[section]

\begin{document}

\begin{frontmatter}

\title{Consensus conditions of continuous-time multi-agent systems with time-delays and measurement noises\thanksref{footnoteinfo}} 

\thanks[footnoteinfo]{This paper was not presented at any IFAC
meeting. Corresponding author: Tao Li. Phone: +86-21-54342646. Fax: +86-21-54342609.}

\author[xf1,xf2]{Xiaofeng Zong}\ead{zongxf8@cug.edu.cn},    
\author[lt]{Tao Li}\ead{tli@math.ecnu.edu.cn},               
\author[jf1,jf2]{Ji-Feng Zhang}\ead{jif@iss.ac.cn}  

\address[xf1]{ School of Automation, China University of Geosciences, Wuhan 430074, China.}  
\address[xf2]{ Hubei Key Laboratory of Advanced Control and Intelligent Automation for Complex Systems, Wuhan 430074, China.}
\address[lt]{Department of Mathematics, East China Normal University, Shanghai 200241, China.}             
\address[jf1]{Key Laboratory of Systems and Control, Academy of Mathematics and Systems Science, Chinese Academy of Sciences, Beijing 100190,  China}
\address[jf2]{School of Mathematical Sciences, University of Chinese Academy of Sciences, Beijing 100049,  China}     


\begin{abstract}                          
This work is concerned with stochastic consensus conditions of multi-agent systems with both time-delays and measurement noises. For the case of additive noises, we develop some necessary conditions and sufficient conditions for stochastic weak consensus by estimating the differential resolvent function for delay equations. By the martingale convergence theorem, we obtain  necessary conditions and  sufficient conditions for stochastic strong consensus. For the case of multiplicative noises, we consider two kinds of time-delays, appeared in the measurement term and the noise term, respectively. We first show that stochastic weak consensus with the exponential convergence rate implies stochastic strong consensus.  Then by constructing degenerate Lyapunov functional, we find the sufficient consensus conditions and show that stochastic  consensus can be achieved by carefully choosing the control gain according to the noise intensities and the time-delay in the measurement term.
\end{abstract}

\end{frontmatter}

\section{Introduction}
The research on consensus in multi-agent systems, which involves coordination of multiple entities with only limited neighborhood information to reach a global goal for the entire team, has offered promising support for solutions in distributed systems, such as flocking behavior and swarms  \citep*{LPP2003,MGFJ2014,ZXHMT2017}, sensor networks \citep*{ASSC2002,OFL2004}. Due to that time-delays are unavoidable in almost all practical systems and the real communication processes are often disturbed by various random factors, each agent cannot measure its neighbors' states timely and accurately. Hence, there has been substantial and increasing interest in recent years in the consensus problem of the multi-agent systems subject to the phenomenon of time-delay and  measurement noise (or stochasticity).

So far lots of achievements have been made in the research of consensus problems of multi-agent systems with  time-delays.   
\cite{Olfati-Saber2004}  presented that small time-delay does not affect the consensus property of the protocol. 
\cite{LR2014IEEEAC} studied a constrained consensus problem for multi-agent systems in unbalanced networks in the presence of time-delays. For the case of distributed  time-delays, 
\citet*{MPA2011} showed that the consensus for single integrator multi-agent systems can be reached under the same conditions as the delay-free case. For the high-order linear multi-agent systems, the  time-delay bound was investigated in  \citet*{CO2011,WZFZ2017}. The mentioned papers above are for the continuous-time models. For the discrete-time models,  we refer to \citet*{SLX2011SICON,HC2014,SN2013} and the references therein.

 Additive and multiplicative noises have been used to model the measurement uncertainties in multi-agent systems. Different from the deterministic consensus dynamics, in the presence of noises, the convergence of stochastic consensus dynamics presents various kinds of probabilistic meanings, where the almost sure consensus and the mean square consensus are of the most practical interest. Note that mean square convergence and almost sure convergence cannot generally imply each other  (see \cite{Mao2007}). So analyzing the relationship between the two kinds of stochastic consensus is imperative and important. To date,  many literatures have been devoted to stochastic consensus analysis of multi-agent systems with measurement noises.

  Additive noises in multi-agent systems are often considered as external interferences and independent of agents' states.  For discrete-time models, distributed stochastic approximation method was introduced for multi-agent systems with additive noises, and  mean square and  almost sure consensus conditions were obtained in 
  \citet*{HM2009,KM2009TSP,LZ2010TAC,HDNM2010Automatica,XZX2012TAC,AB2010}.
For continuous-time models, the necessary and sufficient conditions of mean square average-consensus were obtained in \citet*{LZAutomatica2007,CHTW2011TAC}.
And the sufficient conditions of almost sure strong consensus were stated in \cite{WZ2009}.
\cite{TL2015} gave the relationship between the convergence rate of the consensus error and a representative class of consensus gains in both mean square and probability one.

  Multiplicative noises can be generated by data transmission channels, both during the propagation of radio signals and under signal processing by receivers or detectors. Multiplicative noises have been investigated intensively in \cite{T2002book}
  for signal processing. In multi-agent systems, multiplicative noises have to be considered when there is channel fading or logarithmic quantization \citep*{CFSZ2008,DJ2010,WE2013,LWZ2014TAC}.
  For the multi-agent systems with multiplicative noises,  
  \citet*{NL2013SCL}  investigated the consensus problems of continuous-time systems with the noise intensities being proportional to the absolute value of the relative states of agents. Then this work was extended to the discrete-time version in \citet*{LLX2014RNC}.
   \citet*{LWZ2014TAC} studied the distributed averaging with general multiplicative noises and developed some necessary conditions and sufficient conditions for mean square and almost sure average-consensus. Taking the two classes of measurement noises into consideration,  
   \citet*{ZLZ2016}  gave the necessary and sufficient conditions of mean square and almost sure weak and strong consensus for continuous-time models.

When time-delays and noises coexist in real multi-agent networks, these works above are far from enough to deal with the consensus problem. Based on this phenomenon, the distributed consensus problem was addressed in \citet*{LXZ2011AA}, 
 and  the approximate mean square consensus problem was examined in \citet*{AFJV2015}
 for discrete-time models. For continuous-time models,
 \citet*{LLXZ2011Automatica} presented some sufficient conditions for the mean square average-consensus.  However, the mean square weak consensus,  and  the almost sure  consensus  have not been taken into account even for the case with balanced graphs. Moreover, the works stated above are for the additive measurement noise and little is known about the consensus conditions for the case with the  noises coupled with the delayed states (multiplicative noises).

Motivated by the above discussions and partly based on our recent works \citet*{ZLZ2016,LWZ2014TAC,LZAutomatica2007}, this work investigates the distributed consensus problem of continuous-time multi-agent systems with time-delays and measurement noises, including the additive and multiplicative cases.   Due to the presence of noises, existing techniques for the case with only time-delay  \citep*{Olfati-Saber2004,XZX2013AA} are no longer applicable to the analysis of stochastic consensus.  Moreover,  the coexistence of time-delays and noises leads to the difficulty in finding the relationship between control parameters and time-delays for stochastic consensus problem. Note that even for the case with uniform time-delays, the consensus analysis is not easy due to the presence of noises.  In this paper,  the differential resolvent
function and degenerate Lyapunov functional methods are developed to overcome the difficulties induced by time-delays and noises.

  We first use a variable transformation to transform the closed-loop system into a stochastic differential delay equation (SDDE) driven by the additive or multiplicative noises. Then the key is to analyze the asymptotic stability of SDDEs. Hence, our concern is not only important in the consensus analysis mentioned above but also has its own mathematical interest because the relevant stochastic stability theory for this kind of SDDEs has not been well established. By semi-decoupling the corresponding SDDEs, using differential resolvent function and degenerate Lyapunov functional methods for stability analysis, stochastic consensus problem is solved.  The contribution of the current work can be concluded as  follows.

1) Additive noises case:  We established some new explicit necessary conditions and sufficient conditions for various stochastic consensus under general digraphs.
       \begin{itemize}
         \item For weak consensus, we show that if the digraph contains a spanning tree, then for   any fixed time-delay $\tau_1$ and noise intensity, (a)  mean square weak consensus can be achieved by designing control gain function $c(t)$ satisfying 
             $\int_0^{\infty}c(t)dt=\infty$  and $\lim_{t\rightarrow\infty} c(t)=0$; (b)  almost sure weak consensus can be achieved by designing  control gain function $c(t)$ satisfying 
             $\int_0^{\infty}c(t)dt=\infty$ and $\lim_{t\rightarrow\infty}c(t)\log\int_0^tc(s)ds=0$;
         \item   For strong consensus, we show that if the digraph contains a spanning tree, then for   any fixed time-delay $\tau_1$ and noise intensity, mean square and almost sure strong consensus can be achieved by designing control gain function $c(t)$ satisfying $\tau_1  \max_{2\leq j\leq N}\frac{|\lambda_j|^2}{Re(\lambda_j)}$ $\sup_{t\geq t_0} c(t)<1$ for certain $t_0\geq0$, $\int_0^{\infty}c(t)dt=\infty$ and $\int_0^\infty c(t)^2dt<\infty$, where   $\{\lambda_i\}_{2\leq i\leq N}$ are the non-zero eigenvalues of the corresponding  Laplacian matrix. The mean square strong consensus results relax the restriction of balanced graph and time-delay bound  in \cite{LLXZ2011Automatica}.
       \end{itemize}

2) Multiplicative noises case: We first develop a fundamental theorem to show that mean square (or almost sure) weak consensus with the exponential convergence rate implies mean square (or almost sure) strong consensus. Then by constructing a degenerate  Lyapunov functional, we prove that if the graph is strongly connected and undirected, then for any fixed time-delay $\tau_1$ in the  deterministic measurement and  noise intensity bound $\bar{\sigma}$,  mean square and almost sure strong consensus can be achieved by designing  control gain $k\in( 0,\frac{1}{\lambda_N\tau_1+\frac{N-1}{N} \bar{\sigma}^2})$.

3) The new findings: (1) Mean square weak consensus may not imply almost sure weak consensus, and stochastic weak consensus may not imply stochastic strong consensus  for the case with additive noises; (2) Mean square weak consensus with the exponential  convergence rate implies almost sure strong consensus for the case with multiplicative noises, and stochastic consensus does not necessarily depend on the time-delay in the noise term.

 The rest of the paper is organized as follows. Section \ref{S2} serves as an introduction to the networked systems and consensus problems. Section  \ref{S3add} gives some necessary conditions and  sufficient conditions of stochastic weak and strong consensus for multi-agent systems with time-delay and additive noises.  Section  \ref{S4multi}  aims to consider stochastic consensus problem of multi-agent systems with time-delays and multiplicative noises. 
 Section  \ref{S6con} gives some concluding remarks and discusses the future research topics.

\textbf{Notations:}\quad 
For any complex number $\lambda$ in complex space $\mathbb{C}$, $Re(\lambda)$ and $Im(\lambda)$ denote its real and imaginary parts, respectively, and $|\lambda|$ denotes its modulus. $\mathbf{1}_n$ denotes a $n$-dimensional column vector with all ones. $\eta_{N,i}$ denotes the $N$-dimensional column vector with the $i$th element being $1$ and others being zero.   $I_N$ denotes the $N$-dimensional identity matrix. For a given matrix or vector $A$, its transpose is denoted by $A^T$, and its Euclidean norm is denoted by $\|A\|$.  For two matrices $A$ and $B$, $A\otimes B$ denotes their Kronecker product. For $a,b\in\mathbb{R}$, $a\vee b=\max\{a,b\}$ and $a\wedge b=\min\{a, b\}$. For any given real symmetric matrix $K$, we denote its maximum and minimum eigenvalues by $\lambda_{\max}(K)$ and $\lambda_{\min}(K)$, respectively. Let $(\mathbf{\Omega}, {\mathcal{F}}, \mathbb{P})$ be a complete probability space with a filtration $\{{\mathcal{F}}_t\}_{t\geq 0}$ satisfying the usual conditions.
For a given random variable or vector $X$, its mathematical expectation is denoted by $\mathbb{E}X$. For a (local) continuous martingale $M(t)$, its quadratic variation is denoted by  $\langle M\rangle(t)$. (see \cite{RY1999}).
For  $\tau>0$,  $C ([-\tau,0];\mathbb{R}^n)$ denotes the space of all continuous $\mathbb{R}^n$-valued functions $\varphi$ defined on $[-\tau,0]$.

\section{Problem formulation}\label{S2}
Consider $N$ agents distributed according to  a digraph $\mathcal{G}=\{\mathcal{V}, \mathcal{E},\mathcal{A}\}$, where $\mathcal{V}=\{1,2,...,N\}$ is the set of nodes with $i$ representing the $i$th agent,  $\mathcal{E}$ denotes the set of edges and  $\mathcal{A}$$=$$[a_{ij}]$$\in$$\mathbb{R}^{N\times N}$ is the adjacency matrix of $\mathcal{G}$ with element $a_{ij}=1$ or $0$ indicating whether or not there is an information flow from agent $j$ to agent $i$ directly.  $N_{i}$ denotes the set of the node $i$'s neighbors, that is, $a_{ij}=1$ for $j\in N_{i}$. Also, $\mbox{deg}_{i}=\sum_{j=1}^{N}a_{ij}$ is called the degree of $i$. The Laplacian matrix of $\mathcal{G}$ is defined as $\mathcal{L}=\mathcal{D}-\mathcal{A}$, where $\mathcal{D}=\mbox{diag}(\mbox{deg}_{1},...,\mbox{deg}_{N})$. If $\mathcal{G}$ is balanced, then $\widehat{\mathcal{L}}=\frac{\mathcal{L}^T+\mathcal{L}}{2}$ denotes the Laplacian matrix of the mirror digraph $\widehat{\mathcal{G}}$ of $\mathcal{G}$ (\cite{Olfati-Saber2004}).

For agent $i$, denote its state at time $t$ by $x_i(t)\in\mathbb{R}^n$. In real multi-agent networks, for each agent, the information from its neighbors may have time-delays and noises.  Hence, we consider that the state of each agent is updated by the rule
\begin{equation}\label{mainagent}
\dot{x}_{i}(t)=K(t)\sum_{j=1}^{N}a_{ij}z_{ji}(t),\ i=1,2,...,N, t>0,
\end{equation}
with
\begin{equation}\label{inf1}
z_{ji}(t)=\Delta_{ji}(t-\tau_1) +f_{ji}(\Delta_{ji}(t-\tau_2))\xi_{ji}(t)
\end{equation}
 denoting the measurement of relative states by agent $i$ from its neighbor $j\in N_i$. Here,  $\Delta_{ji}(t)=x_j(t)-x_{i}(t)$, $K(t)\in\mathbb{R}^{n\times n}$ is the control gain matrix function to be designed, $\tau_1\geq0$ and $\tau_2\geq0$ are  time-delays, $\xi_{ji}(t)\in\mathbb{R}$ denotes the measurement noise and $f_{ji}:\mathbb{R}^{n}\mapsto\mathbb{R}^{n}$ is the intensity function. Let $\tau=\tau_1\vee\tau_2$ and the initial data $x_i(t)=\psi_i(t) $ for $t\in[-\tau,0]$, $i=1,2,\cdots,N$ be deterministic continuous functions.  Let $x(t)=[x_1^T(t),...,x_{N}^T(t)]^{T}$ and $\psi(t)=[\psi_1^T(t),\ldots, \psi_N^T(t)]^T$.

 In this work, $\Delta_{ji}(t-\tau_1)$ in \eqref{inf1} is called the measurement term and $f_{ji}(\Delta_{ji}(t-\tau_2))\xi_{ji}(t)$ is called the noise term.   We also assume that the measurement noises are independent Gaussian white noises. In fact, the Gaussian white noise  is a classical assumption in continuous-time models and has been discussed in
\cite{T2002book} for signal processing due to some physical and statistic characteristics. Here, the independence assumption would be conservative, however, to reduce this conservatism with serious mathematical analysis would need more efforts in future investigation.

\begin{assumption}\label{Assum-noise}
 The noise process  $\xi_{ji}(t)\in\mathbb{R}$ satisfies $\int_{0}^{t}\xi_{ji}(s)ds=w_{ji}(t)$, $t\geq0, j,i=1,2,\ldots, N$, where $\{w_{ji}(t), i,j=1,2,...,N\}$ are independent Brownian motions.
\end{assumption}

Note that the noise term in \eqref{inf1} includes the  two cases: First,  the noises in \eqref{inf1} are additive, that is, each intensity $f_{ji}(\cdot)$ is independent of the  agents' states; Second, the noises are multiplicative, that is, the intensity $f_{ji}(\cdot)$  depends on the relative states.
Then the key in stochastic consensus problem is to find an appropriate control gain function $K(t)$ such that the agents reach  mean square or almost sure consensus under the two types of noises.

 \begin{remark}
 \rm{
 Time-delay, multiplicative and additive noises often exist in measurements and information transmission (see \cite{T2002book}).  \cite{Olfati-Saber2004} studied the continuous-time consensus with the measurement delay. 
 \cite{WE2013} and
  \cite{LWZ2014TAC} considered the noisy and delay-free measurement $z_{ji}(t)=x_j(t)-x_i(t)+f_{ji}(x_j(t)-x_i(t))\xi_{ji}(t)$ for the discrete-time and continuous-time models, respectively. The measurement model \eqref{inf1} is the generalization of the noisy measurement model in \cite{LWZ2014TAC}  and the delayed measurement model in   \citet*{Olfati-Saber2004}.
Generally, the ideal measurement $x_j(t)-x_i(t)$ cannot be obtained accurately and timely due to measurement noises and delays. There are measurement delay $\tau_1$ and time-delay $\tau_2$ for the impact of agents' states on the noise intensities.
 Here, the term $f_{ji}(x_j(t-\tau_2)-x_i(t-\tau_2))\xi_{ji}(t)$ can be considered as the  joint impact of time-delay and measurement noises on the ideal measurement $x_j(t)-x_i(t)$.
  }
 \end{remark}

 Here, the two consensus definitions are given as follows.

\begin{definition} \label{defi1}
 The agents are said to reach mean square weak consensus if the system (\ref{mainagent}) with (\ref{inf1})  has the property that for any initial data  $\psi \in C([-\tau,0],\mathbb{R}^{Nn})$ and all distinct $i,j\in \mathcal{V}$, $\lim_{t\rightarrow\infty}\mathbb{E}\|x_i(t)-x_j(t)\|^2=0$.  If, in addition,  there is a random vector $x^{*}\in\mathbb{R}^{n}$, such that $\mathbb{E}\|x^{*}\|^2<\infty$ and $\lim_{t\rightarrow\infty}\mathbb{E}\|x_{i}(t)-x^{*}\|^2=0$, $i=1,2,...,N$, then the agents are said to reach mean square strong consensus. Particularly, if $\mathbb{E}x^{*}=\frac{1}{N}\sum_{j=1}^{N}x_j(0)$, then the agents are said to reach asymptotically unbiased mean square average-consensus (AUMSAC).
\end{definition}

\begin{definition}\label{defi2}
The agents are said to reach almost sure weak consensus if the system (\ref{mainagent}) with (\ref{inf1}) has the property that for any  initial data  $\psi \in C([-\tau,0],\mathbb{R}^{Nn})$ and all distinct $i,j\in \mathcal{V}$, $\lim_{t\rightarrow\infty}\|x_i(t)-x_j(t)\|=0$ almost surely (a.s.) or in probability one. If, in addition, there is a random vector $x^{*}\in\mathbb{R}^{n}$, such that $\mathbb{P}\{\|x^{*}\|<\infty\}=1$ and $\lim_{t\rightarrow\infty} \|x_{i}(t)-x^{*}\| =0$, a.s. $i=1,2,...,N$, then the agents are said to reach almost sure strong consensus. Particularly, if  $\mathbb{E}x^{*}=\frac{1}{N}\sum_{j=1}^{N}x_j(0)$, then the agents are said to reach  asymptotically unbiased  almost sure average-consensus (AUASAC).
\end{definition}

\begin{remark}
  \rm{Definition \ref{defi2} follows that in  \cite{TJ2008} and we use the almost sure consensus to denote such asymptotical behavior.
  Most existing literature on stochastic multi-agent systems with noises and time-delay focused on the mean square consensus. However, in many applications, the result in the sense of probability one is much more reasonable since people can only observe the trajectory of the networks in one random experiment.
  Note that  almost sure convergence and mean square convergence may not imply each other in stochastic systems (see \cite{Mao2007}). Generally, the analysis of mean square convergence is easier than that of almost sure convergence since taking mean square yields a deterministic system.  }
\end{remark}

 We first introduce the following auxiliary lemma (see \citet{ZLZ2016}).
  \begin{lemma}\label{lemma1}
 For the Laplacian matrix $\mathcal{L}$, we have the following assertions:
  \begin{enumerate}
  \item There exists a probability measure $\pi$ such that $\pi^T \mathcal{L}=0$.
    \item  There exists a matrix $\widetilde{Q}\in\mathbb{R}^{N\times(N-1)}$ such that the matrix $Q=(\frac{1}{\sqrt{N}}\mathbf{1}_{N},\widetilde{Q})\in\mathbb{R}^{N\times N}$ is nonsingular and
    \begin{equation}  \label{equationlem}
Q^{-1}=\left(
  \begin{array}{c}
    \nu^T \\
   \overline{Q} \\
  \end{array}
\right),
Q^{-1} \mathcal{L} Q=\left(
  \begin{array}{cc}
    0 & 0\\
   0 &  \widetilde{\mathcal{L}} \\
  \end{array}
\right),
\end{equation}
 where $\overline{Q}\in\mathbb{R}^{(N-1)\times N}$, $\widetilde{\mathcal{L}}\in\mathbb{R}^{(N-1)\times(N-1)}$ and $\nu$ is a left eigenvector of $\mathcal{L}$ such that $\nu^T\mathcal{L}=0$ and  $\frac{1}{\sqrt{N}}\nu^T\mathbf{1}_{N}=1$.
  \item The digraph $\mathcal{G}$ contains a spanning tree if and only if  each eigenvalue of $\widetilde{\mathcal{L}}$ has positive real part. Moreover, if the digraph $\mathcal{G}$ contains a spanning tree, then the probability measure $\pi$ is unique and $\nu=\sqrt{N}\pi$.
 \end{enumerate}
 Especially, if the digraph is balanced, then $\pi=\frac{1}{N}\mathbf{1}_N$ and $Q$ can be constructed as an orthogonal matrix with the form $Q=(\frac{1}{\sqrt{N}}\mathbf{1}_{N},\widetilde{Q})$ and the inverse of $Q$ may be represented in the form $Q^{-1}=\Big[\begin{array}{c}
   \frac{1}{\sqrt{N}}\mathbf{1}_N^T \\
  \widetilde{Q}^T
  \end{array}\Big].$
 \end{lemma}

\section{Networks with time-delay and additive noises}\label{S3add}
In this section, we consider the case with additive noises,  which is concluded as the following assumption.
 \begin{assumption}\label{Assumonf-additive}
For any $x\in\mathbb{R}^n$, $f_{ji}(x)=\sigma_{ji}\mathbf{1}_n$  with $\sigma_{ji}>0$, $i,j=1,\ldots, N$.
\end{assumption}
This assumption has been examined in \citet{AFJV2015} and \cite{HM2009}
for the discrete-time models, and in \cite{LZAutomatica2007}
 for the continuous-time models. Note that under Assumption \ref{Assumonf-additive}, time-delay $\tau_2$ vanishes in the network system.  For the case with additive noises, we choose  $K(t)=c(t)I_n $, where $c(t)\in C((0,\infty);[0,\infty))$. Define $\bar{c}_{t_0}:=\sup_{t\geq t_0} c(t)$, $t_0\geq0$. In fact, the following conditions on the control gain function $c(t)$ were addressed before:
\begin{description}
  \item[(C1)] $\int_0^{\infty}c(t)dt=\infty $;
  \item[(C2)] $\int_{0}^{\infty}c^2(t)dt<\infty$;
  \item[(C3)] $\lim_{t\rightarrow\infty}c(t)=0$.
\end{description}

\begin{remark}
   \rm{Conditions (C1) and (C2) are called convergence condition and  robustness condition, respectively \citep*{LZAutomatica2007}. In fact, the two conditions can be regarded as the continuous-time version of the classical rule for the step size in discrete-time stochastic approximation, which intuitively means that the decay of gain function is allowed, but cannot be too fast.
}
\end{remark}

For the systems with additive noises, the necessary and sufficient conditions of  mean square and  almost sure strong and weak consensus seems to be clear now in view of \cite{ZLZ2016}.
When time-delay appears, the sufficient conditions involving (C1) and (C2) were obtained for mean square strong consensus in \cite{LLXZ2011Automatica}
 under  balanced graphs. But little is known about the necessary and sufficient conditions of stochastic strong and weak consensus under general digraphs. This section will fill in this gap.

Here, we first consider the linear scalar equation
\begin{equation}\label{SDE04091}
  \dot{\bar{X}}(t)=-\lambda c(t)\bar{X}(t-\tau_1), t>0,
\end{equation}
$\bar{X}(t)=\xi(t)$ for $t\in[- \tau_1 ,0]$, where $Re(\lambda)>0$, $ \tau_1 \geq0$ and $\xi\in C ([- \tau_1 ,0],\mathbb{C})$. The solution to \eqref{SDE04091} has the form  \citep*{GLS1990}
$
  \bar{X}(t)= \Gamma(t,s)\bar{X}(s), \ \forall \ t\geq s\geq0,
 $
where $\Gamma(t,s)$ is the differential resolvent function, satisfying $\Gamma(t,t)=1$ for $t>0$, $\Gamma(t,s)=0$ for $t<s$ and
\begin{equation}\label{resolve0}
\frac{\partial}{\partial t}\Gamma(t,s)=-\lambda c(t) \Gamma(t- \tau_1,s), t>s.
\end{equation}

Although some papers have studied the asymptotic stability of the linear equation \eqref{SDE04091} (see \cite{GY1972}, \cite{HL1993} for example),
the decay rate has not been revealed. The following lemma is to estimate the decay rate of differential  resolvent function $\Gamma(t,s)$. The proof is given in Appendix.
\begin{lemma}\label{lemmart}
    If  there is a constant $t_0\geq0$ such that $\tau_1 \bar{c}_{t_0}\frac{|\lambda|^2}{Re(\lambda)}<1$, then the solution to \eqref{resolve0} satisfies
   \begin{equation}\label{estimater0}
|\Gamma(t,s)|^2\leq b(\lambda)  e^{-\varrho(\lambda)\int_s^tc(u)du}, \ t>s\geq t_0.
\end{equation}
Here,  $b(\lambda)$ is a positive constant depending on $\lambda$ and $\varrho(\lambda):=\rho_1(\lambda)\wedge\rho_2(\lambda)$, where $\rho_1(\lambda)$ is the unique  root of the equation $3\rho|\lambda|^2 \tau_1 ^2 \bar{c}_{t_0}^2 e^{\rho\bar{c}_{t_0} \tau_1 }+2\rho-2(Re(\lambda)-|\lambda|^2 \tau_1 \bar{c}_{t_0})=0$ and $\rho_2(\lambda)=\frac{1}{\bar{c}_{t_0} \tau_1 }\log\frac{1}{|\lambda| \bar{c}_{t_0}\tau_1 }$. 
\end{lemma}
\begin{remark}
 \rm{ Due to the time-delay, we cannot use the similar methods in \cite{ZLZ2016} to obtain the mean square and almost sure consensus conditions since we do not have the explicit expression of $\Gamma(t,s)$. However, we can have  the decay rate estimation of $\Gamma(t,s)$, which is established in Lemma \ref{lemmart} and plays an important role in obtaining the  sufficient conditions for mean square and almost sure  consensus. 
 }
\end{remark}
  By Lemma \ref{lemmart}, we now examine mean square and  almost sure consensus, respectively.

\subsection{Mean square consensus}
Let $\varrho(\lambda)$ be defined in Lemma \ref{lemmart} and $\{\lambda_i\}_{i=2}^N$ be the  eigenvalues of $\widetilde{\mathcal{L}}$. Define $\varrho_0=\min_{1\leq j\leq N} \varrho(\lambda_{j})$ and $\bar{\lambda}=\max_{2\leq i\leq N}Re(\lambda_i(\mathcal{L}))$. We introduce another conditions on the control gain $c(t)$:
\begin{description}
  \item[(C4)] $\lim_{t\rightarrow\infty}\int_0^t e^{-\varrho_0\int_s^tc(u)du}c^2(s)ds=0$;
  \item[(C4$'$)]   $\lim_{t\rightarrow\infty}\int_0^t e^{-2\bar{\lambda}\int_s^tc(u)du}c^2(s)ds=0$.
\end{description}

\begin{remark}
   \rm{At the first glance, (C4) and (C4$'$) are very complicated, in fact, they correspond to the sufficient condition and necessary condition  for mean square stability of SDEs with additive noises in \citet{ZLZ2016}.
 Moreover, thanks to (C4) and (C4$'$), we can find much simpler conditions for mean square weak consensus (see Corollary \ref{coro1} and Remark \ref{remark3.2} below).}
\end{remark}

\begin{theorem}\label{thadddelay}
For system (\ref{mainagent}) with (\ref{inf1}) and $K(t)=c(t)I_{n}$, suppose that Assumptions \ref{Assum-noise} and \ref{Assumonf-additive} hold, and $\tau_1 \bar{c}_{t_0}\max_{2\leq j\leq N}\frac{|\lambda_j|^2}{Re(\lambda_j)}<1$ for certain $t_0\geq0$. Then the agents reach mean square weak  consensus if  $\mathcal{G}$ contains a spanning
tree and conditions (C1) and (C4) hold,  and only if  $\mathcal{G}$ contains a spanning
tree and condition (C4$'$) holds under (C1).
\end{theorem}

\textbf{Proof}
 Substituting  (\ref{inf1}) into (\ref{mainagent}) and using Assumption \ref{Assumonf-additive} produce
$
dx(t)=-c(t)(\mathcal{L}\otimes I_n)x(t-\tau_1)dt+c(t)\sum_{i,j=1}^{N}a_{ij}\sigma_{ji}(\eta_{N,i}\otimes \mathbf{1}_n)dw_{ji}(t).
$
Let $\nu$ be defined in Lemma \ref{lemma1} and $J_{N}=\frac{1}{\sqrt{N}}\mathbf{1}_N\nu^{T}$.
Noting that $\mathcal{L}\mathbf{1}_{N}=0$ and $\nu^T\mathcal{L}=0$, then $(I_N-J_N)\mathcal{L}= \mathcal{L}(I_N-J_N)$. Let $\delta(t)=[(I_N-J_N)\otimes I_n]x(t)=[\delta_1^T(t),...,\delta_N^T(t)]^{T}$, where $\delta_{i}(t)\in\mathbb{R}^{n}$, $i=1,2,...,N$. Then we have
$
d\delta(t) = -c(t)(\mathcal{L}\otimes I_n)\delta(t-\tau_1)dt +c(t)\sum_{i,j=1}^{N}a_{ij}\sigma_{ji}((I_N-J_N)\eta_{N,i}\otimes \mathbf{1}_n)dw_{ji}(t).
$
 Define $\widetilde{\delta}(t)=(Q^{-1}\otimes I_n)\delta(t)=[\widetilde{\delta}_1^T(t),\ldots,\widetilde{\delta}_N^T(t)]^T$, $\overline{ \delta}(t)=[\widetilde{\delta}_2^T(t),\ldots,\widetilde{\delta}_N^T(t)]^T$, $\widetilde{\delta}_i(t)\in\mathbb{R}^n$. By the definition of $Q^{-1}$ given in Lemma \ref{lemma1}, we have $\widetilde{\delta}_1(t) =(\nu^T\otimes I_n)\delta(t)=(\nu^T(I_N-J_N)\otimes I_n)x(t) =0$ and
\begin{eqnarray}\label{asd5128}
  \quad d\overline{\delta}(t)=-c(t)(\widetilde{\mathcal{L}}\otimes I_n)\overline{\delta}(t-\tau_1)dt + dM(t),
\end{eqnarray}
where $\overline{Q}$ is defined in Lemma \ref{lemma1} and $M(t)= \sum_{i,j=1}^{N}a_{ij}\sigma_{ji}(\bar{q}_i\otimes \mathbf{1}_n)\int_0^tc(s)dw_{ji}(s)$, and $\bar{q}_i=\overline{Q}(I_N-J_N)\eta_{N,i}$. Note that $\delta_i(t)=x_i-\frac{1}{\sqrt{N}}\sum_{k=1}^N\nu_kx_k(t)=\frac{1}{\sqrt{N}}\sum_{k=1}^N\nu_k(x_i-x_k)$ and then $x_j(t)-x_i(t)=\delta_j(t)-\delta_i(t)$. Hence,  mean square weak consensus equals $\lim_{t\to\infty}\mathbb{E}\|\overline{\delta}(t)\|^2=0$ for any initial data.  By the matrix theorem, there exists a complex invertible matrix $R$ such that $R\widetilde{\mathcal{L}}R^{-1}=J$, Here, $J$ is  the Jordan normal form of $\widetilde{\mathcal{L}}$, i.e., $J=diag(J_{\lambda_2,n_2},\ldots,J_{\lambda_l,n_l}),\ \sum_{k=2}^ln_k=N-1,$
where $\lambda_2,\lambda_3,\ldots,\lambda_l$ are all
the eigenvalues of $\widetilde{\mathcal{L}}$ and
$
J_{\lambda_k,n_k}
$
is  the corresponding Jordan block of size $n_k$ with eigenvalue $\lambda_k$.   Letting $ Y(t)=(R\otimes I_n)\bar{\delta}(t)=[ Y_1^T(t), \ldots,$ $  Y_N^T(t)]^T$ with $ Y_j(t)\in\mathbb{C}^n$, then we have from \eqref{asd5128} that
$
 d Y(t)= -c(t)(J\otimes I_n) Y(t-\tau_1)dt+  (R\otimes I_n)dM(t).
$
 Considering the $k$th Jordan block and its corresponding component $\eta_k(t)=[\eta_{k,1}^T(t),\ldots,\eta_{k,n_k}^T(t)]^T$ and $R(k)=[R_{k,1}^T,\ldots,R_{k,n_k}^T]^T$, where $\eta_{k,j}(t)= Y_{k_j}(t)$ and  $R_{k,j}=R_{k_j}$ is $k_j$th row of $R$ with $k_j=\sum_{l=2}^{k-1}n_l+j$, we have
$
d\eta_k(t)=-c(t)(J_{\lambda_k,n_k}\otimes I_n)\eta_k(t-\tau_1)dt+  (R(k)\otimes I_n)dM(t).
$
This produces the following semi-decoupled delay equations:
 \begin{eqnarray}\label{05190}
   d\eta_{k,n_k}(t)=-c(t)\lambda_k\eta_{k,n_k}(t-\tau_1)dt+ \mathbf{1}_nd M_{k,n_k}(t)
 \end{eqnarray}
and
\begin{eqnarray}\label{05191}
 d\eta_{k,j}(t)&=&-c(t)\lambda_k\eta_{k,j}(t-\tau_1)dt-c(t) \eta_{k,j+1}(t-\tau_1)dt\cr
&&+ \mathbf{1}_ndM_{k,j}(t), \ j=1, \ldots, n_k-1.
 \end{eqnarray}
where $M_{k, j}(t)=\sum_{i=1}^{N}r_{k_j,i}\sum_{j=1}^Na_{ij} \sigma_{ji} \int_0^tc(s)dw_{ji}(s)$, $r_{k_j,i}=R_{k_j}\bar{q}_i$, $j=1, \ldots, n_k$.
Then mean square weak consensus is equivalent to that $\lim_{t\rightarrow\infty}\mathbb{E}\|\eta_{k,j}(t)\|^2=0$, $k=1,\ldots, l$, $j=1,2,\ldots, n_k$ for any initial data $\psi$.

We firstly prove the "if" part. Let $\Gamma_k(t,s)$ denote the differential resolvent function defined by \eqref{resolve0} with $\lambda$ being replaced with $\lambda_k$. Under Assumption \ref{lemma1}, we know that $Re(\lambda_k)>0$ and $\nu=\sqrt{N}\pi$. By means of a variation of constants formula for \eqref{05190}, we obtain
\begin{equation}\label{nk519}
\eta_{k,n_k}(t)=\Gamma_k(t,t_0)\eta_{k,n_k}(t_0) +  \mathbf{1}_nZ_{k,n_k}(t,t_0),
\end{equation}
where $Z_{k,n_k}(t,t_0)=\int_{t_0}^t\Gamma_k(t,s)dM_{k,n_k}(s)$.
Then we get
$
\mathbb{E}\|\eta_{k,n_k}(t)\|^2$ $=|\Gamma_k(t,t_0)|^2 \|\eta_{k,n_k}(t_0)\|^2 + C_{n_k} \int_{t_0}^t|\Gamma_k(t,s)|^2$ $c^2(s)ds,
$
where $C_{n_k}=n\sum_{i=1}^{N}|r_{k_{n_k},i}|^2$ $\sum_{j=1}^Na_{ij} \sigma^2_{ji}$. By Lemma \ref{lemmart}, we have
$
\mathbb{E}\|\eta_{k,n_k}(t)\|^2\leq b(\lambda_k) e^{-\varrho(\lambda_k)\int_{t_0}^tc(u)du}$ $ \|\eta_{k,n_k}(t_0)\|^2+C_{n_k}b(\lambda_k)\int_{t_0}^t c^2(s)$ $e^{-\varrho(\lambda_k)\int_s^tc(u)du}ds.
$ 
 By (C1) and (C4), we have  $\lim_{t\rightarrow\infty}$ $\mathbb{E}\|\eta_{k,n_k}(t)\|^2=0$. Assume that $\lim_{t\rightarrow\infty}\mathbb{E}\|\eta_{k,j+1}(t)\|^2=0$ for some fixed $j<n_k$, and we will show  $\lim_{t\rightarrow\infty}$ $\mathbb{E}\|\eta_{k,j}(t)\|^2=0$.  By means of a variation of constants formula for \eqref{05191}, we obtain
$
\eta_{k,j}(t)=\Gamma_k(t,t_0)\eta_{k,j}(t_0) +  \mathbf{1}_nZ_{k,j}(t)- \int_{t_0}^t\Gamma_k(t,s)c(s) \eta_{k,j+1}(s)ds,
$ where $Z_{k,j}(t)=\int_{t_0}^t\Gamma_k(t,s)dM_{k,j}(s)$.
Hence, we have
$\mathbb{E}\|\eta_{k,j}(t)\|^2$ $\leq 2|\Gamma_k(t,t_0)|^2\mathbb{E}\|\eta_{k,j}(t_0)\|^2+   C_{j} \int_{t_0}^t|\Gamma_k(t,s)|^2c^2(s)ds
 + 2 \mathbb{E}\|\int_{t_0}^t\Gamma_k(t,s)c(s) \eta_{k,j+1}(s)ds\|^2,
$ where $C_j=n\sum_{i=1}^{N}$ $|r_{k_{j},i}|^2\sum_{l=1}^Na_{il} \sigma^2_{li}$.
Note that the first two terms tend to zero, then we only need to prove that  the last term vanishes at infinite time. Let $k,j$ be fixed and write  $\eta_{k,j+1}(s)=[y_1(s),\ldots,y_n(s)]^T\in\mathbb{C}^n$, then $ \lim_{t\rightarrow\infty}\mathbb{E}|y_m(s)|^2=0$, $m=1,\ldots,n$, and
 $
  \mathbb{E}\|\int_{t_0}^t\Gamma_k(t,s)c(s) \eta_{k,j+1}(s)ds\|^2
   \leq b(\lambda_k) \sum_{m=1}^n\mathbb{E}\widetilde{X}^2_m(t),
$ 
where $\widetilde{X}_m(t)=\int_0^te^{-0.5\varrho(\lambda_k)\int_s^tc(u)du}c(s) |y_m(s)|ds$.  By Minkowski's  inequality  for integrals, we have
 $
    \sqrt{\mathbb{E}(\widetilde{X}_m(t))^2} \leq \int_0^te^{-0.5\varrho(\lambda_k)\int_s^tc(u)du}c(s) \sqrt{\mathbb{E}|y_m(s)|^2}ds.
$ Let $U_1(t)=\int_0^t e^{0.5\varrho(\lambda_k)\int_0^sc(u)du}c(s)\sqrt{\mathbb{E}|y_m(s)|^2} ds$.   Then it is easy to see from (C1) that
 $
 \lim_{t\rightarrow\infty}\sqrt{\mathbb{E}\widetilde{X}^2_m(t)} 
$ $ =0
 $ if $\lim_{t\to\infty}U_1(t)<\infty$.
 Note that $\lim_{t\rightarrow\infty}\mathbb{E}|y_m(s)|^2=0$. If $\lim_{t\to\infty}U_1(t)=\infty$, then L'H\^{o}pital's rule gives
$
 \lim_{t\rightarrow\infty}\sqrt{\mathbb{E}(\widetilde{X}_m(t))^2} \leq   \lim_{t\rightarrow\infty}\frac{\sqrt{\mathbb{E}|y_m(t)|^2}}{0.5\varrho(\lambda_k) } =0.
$
 Hence, we have $\lim_{t\rightarrow\infty}\mathbb{E}|\widetilde{X}_m(t)|^2$ $=0$, and then  $\lim_{t\rightarrow\infty}\mathbb{E}\|\eta_{k,j}(t)\|^2=0$ for the fixed $j<n_k$. The similar induction yields $\lim_{t\rightarrow\infty}\mathbb{E}\|\eta_{k,j}(t)\|^2=0$ for all $j=1,\ldots,n_k$, and therefore, $\lim_{t\rightarrow\infty}\mathbb{E}\|\eta_{k,j}(t)\|^2=0$ for all $k=1,\ldots, l$ and  $j=1,\ldots,n_k$. That is, the agents achieve  mean square weak consensus if $\mathcal{G}$ contains a spanning tree and conditions (C1) and (C4) hold.

We now prove the "only if" part. First, if  $\mathcal{G}$ does not contain a spanning tree, then $\mathcal{L}$ at least has two zero eigenvalues. By Lemma \ref{lemma1}, $\widetilde{\mathcal{L}}$~at least has one zero eigenvalue, denoted by $\lambda_2$. Hence, we have from~\eqref{nk519}
 \begin{equation}\label{051901920}
   \eta_{2,n_2}(t)=\eta_{2,n_2}(0)+ \mathbf{1}_n M_{2,n_2}(t),
 \end{equation}
Therefore,
 $
\mathbb{E}\|\eta_{2,n_2}(t)\|^2=\|\eta_{2,n_2}(0)\|^2 + n\mathbb{E}|M_{2,n_2}(t)|^2>0,
 $
which is in contradiction with the definition of mean square weak consensus, that is, ~$\mathcal{G}$~contains a spanning tree.  Second, we need to show the necessity of condition (C4$'$) for mean square weak consensus. Let $\nu=\sqrt{N}\pi$ and  $G_k(t)=\eta_{k,n_k}(t)-\eta_{k,n_k}(t-\tau_1)$, then mean square weak consensus implies  $\lim_{t\rightarrow\infty}\mathbb{E}\|\eta_{k,n_k}(t)\|^2=0$ and $\lim_{t\rightarrow\infty}\mathbb{E}\|G_k(t)\|^2=0$. Note that
$
 d\eta_{k,n_k}(t)=-c(t)\lambda_k\eta_{k,n_k}(t)dt+c(t)\lambda_kG_k(t)dt+ \mathbf{1}_nd M_{k,n_k}(t).
$
By the variation of constants formula, we obtain
\begin{eqnarray}\label{hehe2}
  \eta_{k,n_k}(t)&=&e^{-\lambda_k\int_0^tc(u)du }\eta_{k,n_k}(0)+ \mathbf{1}_nZ_{k,n_k}(t)+U_2(t)\cr
  &=&: \zeta_{k,n_k}(t)+U_2(t).
\end{eqnarray}
where $U_2(t)=\int_0^te^{-\lambda_k\int_s^tc(u)du }c(s)\lambda_kG_k(s)ds$, $\zeta_{k,n_k}$ is the solution to \eqref{05190} with $\tau_1=0$, that is, it satisfies
 \begin{equation}\label{051901101}
   d\zeta_{k,n_k}(t)=-c(t)\lambda_k\zeta_{k,n_k}(t)dt+ \mathbf{1}_nd M_{k,n_k}(t).
 \end{equation}
Then we get
$
  \mathbb{E}\|\zeta_{k,n_k}(t)\|^2\leq 2\mathbb{E}\|U_2(t)\|^2$ $+2\mathbb{E}\|\eta_{k,n_k}(t)\|^2.
$
By the similar methods used in estimating $\mathbb{E}\|\int_{0}^t\Gamma_k(t,s)$ $c(s) \eta_{k,j+1}(s)ds\|^2$ above, we can obtain
$\lim_{t\rightarrow\infty}\mathbb{E}\|U_2(t)\|^2$ $=0,$
 and then $\lim_{t\rightarrow\infty} \mathbb{E}\|\zeta_{k,n_k}(t)\|^2=0$.  It is shown in \citet{ZLZ2016} that
$
\lim_{t\to\infty}$ $\mathbb{E}\|\zeta_{k,n_k}(t)\|^2=0
$
implies condition (C4$'$) under (C1) and $Re(\lambda_k)>0$. Hence, the proof is complete.
$\Box$

It can be seen that Lemma \ref{lemmart} plays an important role in the consensus analysis, where the condition $\tau_1 \bar{c}_{t_0}\max_{2\leq j\leq N}\frac{|\lambda_j|^2}{Re(\lambda_j)}<1$ for certain $t_0\geq0$   is always true   if (C3) holds.  Hence, we can obtain the following corollary. The proof is the same as that in \citet{ZLZ2016} and is omitted.

\begin{corollary}\label{coro1}
    For system (\ref{mainagent}) with (\ref{inf1}) and $K(t)=c(t)I_{n}$, suppose that Assumptions \ref{Assum-noise} and \ref{Assumonf-additive} hold.
    Then the agents achieve mean square weak consensus if  $\mathcal{G}$ contains a spanning tree and conditions (C1) and (C3) hold. Moreover, if $c(t)$ is a decreasing function and satisfies (C1), then the agents achieve  mean square weak consensus  only if  $\mathcal{G}$ contains a spanning tree and   (C3) holds.
\end{corollary}


\begin{remark}\label{remark3.2}
 \rm{ In fact, the proof of  Corollary \ref{coro1} highly depends on Theorem \ref{thadddelay}, where the sufficient condition (C4) and the necessary condition (C4$'$) for mean square weak consensus produce the sufficiency of (C3) and the necessity of (C3) when $c(t)$ is monotonically decreasing, respectively.  Corollary \ref{coro1} is important since it provides the succinct conditions (C1) and (C3), and  implies that condition (C2) is unnecessary for mean square weak consensus. }
\end{remark}


Above, we have obtained the conditions for mean square weak consensus. Now, we can apply the martingale convergence theorem to get the conditions for mean square strong consensus.

\begin{theorem}\label{thadddelay2}
For system (\ref{mainagent}) with (\ref{inf1}) and $K(t)=c(t)I_{n}$, suppose that Assumptions \ref{Assum-noise} and \ref{Assumonf-additive} hold, and $\tau_1 \bar{c}_{t_0}\max_{2\leq j\leq N}\frac{|\lambda_j|^2}{Re(\lambda_j)}<1$ for certain $t_0\geq0$. Then  the agents reach mean square strong consensus if  $\mathcal{G}$ contains a spanning
tree and conditions (C1)-(C2) hold, and only if $\mathcal{G}$ contains a spanning
tree and condition (C2) holds under (C1).
\end{theorem}
\textbf{Proof}
 By the definitions of  mean square weak and strong consensus, we can see that  mean square strong consensus is equivalent to   mean square weak consensus plus that the average $\nu^Tx(t)$ is convergent in the sense of mean square. It is proved in \cite{LZAutomatica2007}  that (C2) under (C1) implies (C4), then from Theorem \ref{thadddelay}, conditions (C1), (C2) and the existence of the spanning tree give  mean square weak consensus. Note that the existence of the spanning tree implies  $\nu=\sqrt{N}\pi\neq0$ and time-delay $\tau_1$ does not change the average of the states of agents, that is,
\begin{equation}
\label{centroidaddadd1}
\nu^T x(t)=\pi^Tx(t)= \pi^Tx(0)+\bar{M}(t),
\end{equation}
where $ \bar{M}(t)= \mathbf{1}_n  \sum_{i,j=1}^{N}a_{ij}\pi_i\sigma_{ji}\int_{0}^{t}c(s) dw_{ji}(s)$. It is easy to see that
 \begin{equation}\label{emt}
    \mathbb{E}\|\bar{M}(t)\|^2=  n\sum_{i,j=1}^{N}a_{ij}^2\pi_i^2\sigma_{ji}^2 \int_0^{t}c^2(s)ds,
 \end{equation}
 and then the mean square convergence of $ \pi^Tx(t)$ is equivalent to the mean square boundedness of $\bar{M}(t)$ \cite[Theorem 1, p.20]{LS1989}, which is guaranteed by (C2).
 Hence, the mean square strong consensus holds with the consensus limit $x^*= \pi^Tx(0)+\bar{M}(\infty)$, where $\bar{M}(\infty):=\lim_{t\rightarrow\infty} \bar{M}(t)$ is a common Gaussian random variable.  If  mean square strong consensus is achieved, then Theorem \ref{thadddelay} implies that $\mathcal{G}$ contains a spanning tree. At the same time, \eqref{emt} and the convergence of $ \pi^Tx(t)$ imply condition (C2). Therefore, the proof is complete.
$\Box$

\begin{remark}\rm{
  Theorem \ref{thadddelay}, Corollary \ref{coro1} and  Theorem \ref{thadddelay2} give the design of control gain for mean square consensus. They show that if  $\mathcal{G}$ contains a spanning tree, then for any given time-delay $\tau_1$,  the control gain function $c(t)$ can be properly designed for guaranteeing  mean square weak and strong consensus. These improve  the results in \cite{LLXZ2011Automatica} in the following three aspects. (a) \citet{LLXZ2011Automatica} considered the case with balanced digraphs, while our consensus analysis is for general digraphs. (b) \citet{LLXZ2011Automatica} require  the time-delay $\tau_1<\frac{\lambda_2(\widehat{\mathcal{L}})}{\|\mathcal{L}\|^2}$, no matter how the control gain functions are selected, while we remove the delay bound restriction and show that for any given time-delay $\tau_1$,  the control gain function can be properly designed for guaranteeing  mean square consensus. (c) Even for the case with $\bar{c}_{0}=1$ and undirected graphs, our delay bound restriction $\lambda_N\tau_1<1$ is weaker than $\lambda_N^2\tau_1<\lambda_2$ in   \citet{LLXZ2011Automatica}. (c) We get not only sufficient conditions for mean square strong consensus, but also the necessary conditions and  sufficient conditions for mean square weak consensus. Here, the main skills are the semi-decoupled method and the differential resolvent function.}
\end{remark}

\subsection{Almost sure consensus}

 Here, we give some necessary conditions and sufficient conditions for almost sure weak and strong consensus. To examine almost sure weak consensus, we need two more conditions:
\begin{description}
  \item[(C5)]  $\lim_{t\rightarrow\infty}c(t)\log\int_0^tc(s)ds=0$;
   \item[(C5$'$)]  $\liminf_{t\rightarrow\infty}c(t)\log\int_0^tc(s)ds=0$.
\end{description}

\begin{remark}
  \rm{ 
Intuitively, (C5) and (C5$'$) mean that the gain function $c(t)$ under (C1) should decay with certain rate and the rate cannot be too large. The two conditions can help us find the fact that mean square weak consensus may not imply almost sure weak consensus.}
\end{remark}

\begin{theorem}\label{thadddelay3}
For system (\ref{mainagent}) with (\ref{inf1}) and $K(t)=c(t)I_{n}$, suppose that Assumptions \ref{Assum-noise},  \ref{Assumonf-additive} and condition (C1) hold.  Then the agents achieve  almost sure weak consensus if  $\mathcal{G}$ contains a spanning tree and condition (C5) holds, and only if  $\mathcal{G}$ contains a spanning tree. Moreover, if $\mathcal{G}$  is undirected, 
 then the agents achieve  almost sure weak consensus only if   $\mathcal{G}$ is connected and condition (C5') holds.
\end{theorem}
 \textbf{Proof}
Note that almost sure weak consensus is equivalent to that for any initial data $\psi$, $\lim_{t\rightarrow\infty} \|\eta_k(t)\|=0$, a.s., $k=1,\ldots, N$. Let $\theta_{k,n_k}(t)=\zeta_{k,n_k}(t)-\eta_{k,n_k}(t)$, where $\zeta_{k,n_k}$ is defined by \eqref{051901101}.
 Then we have
\begin{equation}\label{theta0121}
   \dot{\theta}_{k,n_k}(t)=-c(t)\lambda_k\theta_{k,n_k}(t-\tau_1)+c(t)g_{k,n_k}(t),
\end{equation}
where $g_{k,n_k}(t)=\lambda_k (\zeta_{k,n_k}(t- \tau_1 )-\zeta_{k,n_k}(t))$ is continuous. Noting that  \citet*{ZLZ2016} proved that $\lim_{t\to\infty}\zeta_{k,n_k}(t)=0$ a.s., then we have that $\lim_{t\rightarrow\infty}\|g_{k,n_k}(t)\|=0$,  a.s. By means of a variation of constants formula for equation \eqref{theta0121}, we have $\theta_{k,n_k}(t)=\Gamma_k(t,t_0)\theta_{k,n_k}(t_0)+ \int_{t_0}^t\Gamma_k(t,s)c(s)g_{k,n_k}(s)ds,$
where $\Gamma_{k}(t,s)$ is the differential resolvent function of \eqref{SDE04091} with $\lambda$ being replaced by $\lambda_k$. Let  $b_0=\max_{i=2,\ldots, N}b(\lambda_i)$. Note that (C5) implies $\tau_1 \bar{c}_{t_0}\max_{2\leq j\leq N}$ $\frac{|\lambda_j|^2}{Re(\lambda_j)}<1$ for certain $t_0\geq0$. By \eqref{estimater0}, we get
$
   \|\theta_{k,n_k}(t)\|
   \leq \sqrt{b_0}e^{-0.5\varrho_0\int_{t_0}^tc(u)du}\|\theta_{k,n_k}(t_0)\|+\sqrt{b_0}\int_{0}^te^{-0.5\varrho_0\int_s^tc(u)du}$ $c(s)\|g_{k,n_k}(s)\|ds.
$
Let $p(t)=\int_0^te^{0.5\varrho_0\int_0^sc(u)du}\|g_{k,n_k}(s)\|$ $c(s)ds$ and $\tilde{Y}(t)= p(t)$ $e^{-0.5\varrho_0\int_0^tc(u)du}$, then $p(t)$ is increasing and $\lim_{t\rightarrow\infty}p(t)$ $<\infty$ or $\lim_{t\rightarrow\infty}p(t)=\infty$.
It is easy to see from (C1) that $\lim_{t\rightarrow\infty}\|\tilde{Y}(t)\|=0$ a.s. if $\lim_{t\rightarrow\infty}p(t)<\infty$. But if $\lim_{t\rightarrow\infty}p(t)=\infty$, by  L'H\^{o}pital's rule, we still have
$
 \lim_{t\rightarrow\infty}\|\tilde{Y}(t)\|=\frac{2}{\varrho_0} \lim_{t\rightarrow\infty}\|g(t)\|=0, a.s.
$
 Hence, $\lim_{t\rightarrow\infty}\|\theta_{k,n_k}(t)\|=0$, a.s. This together with $\lim_{t\rightarrow\infty}\|\zeta_{k,n_k}(t)\|=0$ gives $\lim_{t\rightarrow\infty}$ $\|\eta_{k,n_k}(t)\|=0$, a.s.

  We now assume that $\lim_{t\rightarrow\infty}\| \eta_{k,j+1}(t)\|=0$, a.s.  for $j<n_k$, and we will show that $\lim_{t\rightarrow\infty}\| \eta_{k,j}(t)\|=0$, a.s. Let $g_{k,j}(t)=\lambda_k (\zeta_{k,j}(t- \tau_1 )-\zeta_{k,j}(t))$ and $\tilde{g}_{k,j+1}=\zeta_{k,j+1}(t)-\eta_{k,j+1}(t-\tau_1)$, where $\zeta_{k,j}$ is the solution to \eqref{05191} with $\tau_1=0$.  Then we obtain
$
  d\theta_{k,j}(t)=-c(t)\lambda_k\theta_{k,j}(t-\tau_1)dt+c(t)g_{k,j}(t)dt-c(t)\tilde{g}_{k,j+1}(t)dt,
$
which together with the variation of constants formula  implies
$
  \theta_{k,j}(t)=\Gamma_k(t,t_0)\theta_{k,j}(t_0)+ \int_{t_0}^t\Gamma_k(t,s)c(s)g_{k,j}(s)ds- \int_{t_0}^t\Gamma_k(t,s)c(s) \tilde{g}_{k,j+1}(s)ds.
$
Note that \citet{ZLZ2016} proved that $\lim_{t\to\infty}\zeta_{k,j}(t)=0$ a.s. for all $k,j$.    Then we  get $\lim_{t\rightarrow\infty}\|\tilde{g}_{k,j+1}\|=0$, a.s. and $\lim_{t\rightarrow\infty}\|g_{k,j}\|=0$, a.s.
 By the similar skills used in estimating $\|\theta_{k,n_k}(t)\|$, we can obtain $\lim_{t\rightarrow\infty}\|\theta_{k,j}(t)\|=0$, a.s. This together with $\lim_{t\rightarrow\infty}\|\zeta_{k,j}(t)\|=0$ gives $\lim_{t\rightarrow\infty}\|\eta_{k,j}(t)\|=0$, a.s. Hence, almost sure weak consensus follows by mathematical induction. 

 If  almost sure weak consensus is achieved, then $\mathcal{G}$ contains a spanning tree. Otherwise, we have from \eqref{051901920} that in order for $\lim_{t\rightarrow\infty}\eta_{1,n_1}(t)=0$, a.s., the martingale $\mathbf{1}_n M_{1,n_1}(t)$ must converge to   ~$-\eta_{1,n_1}(0)$ for any initial data $\psi$, which is impossible since $\eta_{1,n_1}(0)$ depends on the initial data.

 Next, we show the second assertion. Assume that almost sure weak consensus is achieved, then the existence of a spanning tree is proved above.  If $\mathcal{G}$ is undirected, then all corresponding components of $Y(t)$ have the form \eqref{05190} with $\lambda_k>0$, $k=2,\ldots, N$, $n_k=1$. In order to prove that condition (C5$'$) holds, we only need to show  $\lim_{t\rightarrow\infty}\zeta_{k,n_k}(t)=0$, a.s., since this implies (C5$'$)(see \citet{ZLZ2016}). Note that \eqref{hehe2} implies
 $
 \|\zeta_{k,n_k}(t)\|\leq \|\eta_{k,n_k}(t)\|+\int_0^te^{-\lambda\int_s^tc(u)du}\|G_k(s)\|c(s)ds,
$
   and $\lim_{t\rightarrow\infty}\|G_k(t)\|=0$, a.s. and $  \lim_{t\rightarrow\infty}\eta_{k,n_k}(t)=0$. Then we can use the similar methods in proving  $\lim_{t\rightarrow\infty}\|\tilde{Y}(t)\|=0$ a.s. above to obtain that  $\lim_{t\rightarrow\infty}\|\zeta_{k,n_k}(t)\|=0$. Therefore, condition (C5$'$) holds, and the proof is complete.
$\Box$

\begin{remark}
 \rm{  Based on Corollary \ref{coro1} and Theorem \ref{thadddelay3}, we can see that mean square weak consensus does not imply  almost sure weak consensus. In fact, let $\mathcal{G}$  be strongly connected and undirected, and choose $c(t)= \log^{-1}(4+t)$, which satisfies (C1) and (C3), then  we obtain the  mean square weak consensus form Corollary \ref{coro1}. However, by L'H\^{o}pital's rule, $\lim_{t\rightarrow\infty}c(t)\log\int_0^tc(s)ds$ $=1$, so the  almost sure weak consensus does not hold. }
\end{remark}

The following strong consensus is based on the martingale convergence theorem. 
\begin{theorem}\label{thadddelay33}
For system (\ref{mainagent}) with (\ref{inf1}) and $K(t)=c(t)I_{n}$, suppose that Assumptions \ref{Assum-noise},  \ref{Assumonf-additive} and condition (C1) hold, and $\bar{c}_{t_0}\tau_1 \max_{2\leq j\leq N}$ $\frac{|\lambda_j|^2}{Re(\lambda_j)}<1$ for certain $t_0\geq0$.  Then   the agents achieve  almost sure strong consensus  if and only if  $\mathcal{G}$ contains a spanning tree and condition (C2) holds.
\end{theorem}
\textbf{Proof}
For the "only if" part, the necessity of $\mathcal{G}$ to contain a spanning tree is proved above since almost sure strong consensus implies almost weak consensus. Then we prove the necessity of (C2) under the existence of a spanning tree ($\nu=\sqrt{N}\pi\neq0$). Note that \eqref{centroidaddadd1} holds.   Hence, almost sure strong consensus implies~$\bar{M}(t)$~converges almost surely. This also equals ~$\lim_{t\rightarrow\infty}\langle \bar{M}\rangle (t)<\infty$, a.s., (see~\cite[Proposition 1.8, p. 183]{RY1999}). Note that~$\langle \bar{M}\rangle(t)=n\sum_{i,j=1}^Na^2_{ij}\pi_i$ $\sigma_{ji}^2 \int_0^tc^2(s)ds$. Therefore, almost sure strong consensus implies condition (C2). For the "if" part, we know that if the digraph $\mathcal{G}$ contains a spanning tree and conditions (C1)-(C2) can guarantee~$\lim_{t\rightarrow\infty}\zeta_{k,j}(t)=0$ a.s., $k=2,\ldots, l$, $j=1,2,\ldots, n_k$(see \citet{ZLZ2016}). Using the skills in the proof of the first assertion in Theorem \ref{thadddelay3}, we can easily obtain almost sure weak consensus. Then in order for the almost sure strong consensus, we need to show the almost sure convergence of the martingale $\bar{M}(t)$, which can be guaranteed by condition (C2).
$\Box$


\begin{remark}
 \rm{ Theorems \ref{thadddelay3} and \ref{thadddelay33} give the design of the control gain $c(t)$ for almost sure consensus. In fact,    if $\mathcal{G}$ contains a spanning tree, then for any fixed time-delay $\tau_1$, we can choose the control gain  $c(t)$ satisfying  (C1) and (C5) (or (C2) and $\bar{c}_{t_0}\tau_1 $ $\max_{2\leq j\leq N}\frac{|\lambda_j|^2}{Re(\lambda_j)}<1$ for certain $t_0\geq0$) to ensure  almost sure weak (or strong) consensus. Especially, the gain function $c(t)$ satisfying (C1)-(C3) assures the almost sure strong consensus for any $\tau_1$.
 }
\end{remark}



   Note that   conditions (C2)-(C4) are to attenuate the additive measurement noises. So, if the noises vanish ($\sigma_{ji}=0$), we have the following theorem, which  extends \cite{Olfati-Saber2004}  to the case with digraphs and weakens their delay bound condition
$\tau_1\lambda_N<\frac{\pi}{2}$.
  \begin{theorem}\label{newthe}
   For system (\ref{mainagent}) with (\ref{inf1}) and $K(t)=c(t)I_{n}$, suppose that $\sigma_{ji}=0$, $i,j=1,\ldots, N$, and $\mathcal{G}$ contains a spanning tree. If (C1)  holds and $\bar{c}_{t_0}\tau_1 \max_{2\leq j\leq N}$ $\frac{|\lambda_j|^2}{Re(\lambda_j)}<1$ for certain $t_0\geq0$, then the  agents can reach  the deterministic consensus.
  \end{theorem}


%

\section{Networks with  time-delays and multiplicative noises}\label{S4multi}
In this section, we consider the case with time-delays and multiplicative noises. 
 The following assumption is imposed on the noise intensities.
\begin{assumption}\label{Assumonf}
$f_{ji}(0)=0$ and there exists a constant $\bar{\sigma} \geq 0$ such that for any $x\in\mathbb{R}^n$, $\|f_{ji}(x)\|\leq\bar{\sigma}\|x\|,\ i,j=1,2,...,N$.
\end{assumption}

Assumption \ref{Assumonf} is a general assumption in stochastic systems. 
In fact, the case $f_{ji}(x)=\sigma_{ji} x$ studied in \cite{WE2013} falls in the assumption.   
Based on this assumption, we first have the following lemma.

\begin{lemma}\label{lemconnection}
   For system (\ref{mainagent}) with (\ref{inf1}) and $K(t)=K\in \mathbb{R}^{n\times n}$, suppose  that Assumptions \ref{Assum-noise} and  \ref{Assumonf} hold, and $\mathcal{G}$ contains a spanning tree. If the agents reach mean square (or  almost sure) weak consensus with an exponential convergence rate $\gamma$, that is, $ \mathbb{E}\| x_i(t)-x_j(t)\|^2\leq C e^{-\gamma t}$ (or $\limsup_{t\rightarrow\infty}\frac{\log\| x_i(t)-x_j(t)\|}{t}\leq -\gamma,\ a.s.$ ) for certain $C,\gamma>0$ and any $i\neq j$, then the agents must reach   mean square (or almost sure) strong consensus.
\end{lemma}

Lemma \ref{lemconnection}  tells us that in order to obtain  mean square (or almost sure) strong consensus, we only need to get mean square (or  almost sure) weak consensus with an exponential convergence rate. In the following, we find the appropriate  control gain $K$ such that the agents can achieve  mean square and almost sure consensus.

We will assume that $\mathcal{G}$ is undirected. Then $\nu=\mathbf{1}^T/\sqrt{N}$ and $\widetilde{Q}$ in Lemma \ref{lemma1} can be constructed as $\widetilde{Q}=[\phi_2,...,\phi_{N}]=:\phi$, where $\phi_{i}$ is the unit eigenvector of $\mathcal{L}$ associated with the  eigenvalue $\lambda_i=\lambda_{i}(\mathcal{L})$, that is, $\phi_{i}^{T}\mathcal{L}=\lambda_{i}\phi_{i}^{T}$, $\|\phi_{i}\|=1$, $i=2,...,N$. Hence,  $\widetilde{\mathcal{L}}=\mbox{diag}(\lambda_2, \lambda_3, \cdots, \lambda_N)=:\Lambda$. Continuing to use the definitions of $\delta(t)$ and $\bar{\delta}(t)$ in obtaining \eqref{asd5128} yields
\begin{equation}\label{s4guiyi}
d\overline{\delta}(t)=-(\Lambda \otimes K)\overline{\delta}(t-\tau_1)dt+dM_{\tau_2}(t),
\end{equation}
where $M_{\tau_2}(t)=\sum_{i,j=1}^{N}a_{ij}\int_0^t[\phi^{T}(I_N-J_N)\eta_{N,i}\otimes(Kf_{ji}(\delta_j(s-\tau_2)-\delta_i(s-\tau_2)))]dw_{ji}(s)$. Define the degenerate Lyapunov functional for $\overline{\delta}_t=\{\overline{\delta}(t+\theta):\theta\in[-\tau_1,0]\}$,
\begin{eqnarray}\label{Lfunctional}
 V(\overline{\delta}_t)&=&\int_{-\tau_1}^0\Big[\int_{t+s}^t\overline{\delta}^T(\theta)(\Lambda^2\otimes K^TK)\overline{\delta}(\theta)d\theta\Big]ds\cr
 &&+\|\overline{\delta}(t)-(\Lambda \otimes K)\int_{t-\tau_1}^t\overline{\delta}(s)ds\|^2.
\end{eqnarray}
This is known as degenerate functional in \cite{KM1992}.
Based on \eqref{Lfunctional}, we can get the following theorem.


\begin{theorem}\label{th1}
 For system (\ref{mainagent}) with (\ref{inf1})  and $K(t)=kI_{n}$, suppose that Assumptions \ref{Assum-noise} and  \ref{Assumonf} hold,   and $\mathcal{G}$ is undirected and connected. If
\begin{equation}\label{s4taudefinition}
 0<k<\frac{1}{\lambda_N\tau_1+\frac{N-1}{N} \bar{\sigma}^2},
\end{equation}
then the agents reach  AUMSAC and AUASAC with exponential   convergence rates less than $\gamma_{\tau_2}$ and $\gamma_{\tau_2}/2$ respectively, where $\gamma_{\tau_2}$ is the unique root of the equation
$
2 k(1-\frac{N-1}{N}k\bar{\sigma}^2e^{\gamma \tau_2}-\lambda_Nk\tau_1)\lambda_2-2\gamma-3 \lambda^2_Nk^2\tau_1^2\gamma  e^{\gamma\tau_1} =0.
$ 
Moreover, if $f_{ji}(x)=\sigma_{ij} x$ with $\sigma_{ij}>0,\ i\neq j,\ i, j=1,2,...,N$ and $2\tau_2\geq\tau_1$, then the agents achieve AUMSAC only if   $0<k<\frac{N}{\underline{\sigma}^2(N-1)}$, where $\underline{\sigma}=\min_{i,j=1}^N\sigma_{ji}$.
\end{theorem}
 \textbf{Proof}
Note that $K=kI_n$, $\eta^T_{N, i}(I_N-J_N)\eta_{N, i}=\frac{N-1}{N}, (I_N-J_N)^2=I_N-J_N$ and $\phi\phi^{T}=I_{N}-J_{N}$. Then
$\langle M_{\tau_2}\rangle(t)=\frac{N-1}{N}k^2\sum_{i,j=1}^N a^2_{ij}\int_0^t\|f_{ji}(\delta_j(s-\tau_2)-\delta_i(s-\tau_2))\|^2ds$.
Note that $a_{ij}^2=a_{ij}$, $i,j=1,2,...,N$, and $\delta^T(t)(\mathcal{L}\otimes I_n)\delta(t)=\frac{1}{2}\sum_{i,j=1}^Na_{ij}\|\delta_j(t)-\delta_i(t)\|^2$ \citep*{Olfati-Saber2004}, then from  Assumption \ref{Assumonf}, we obtain
$
\sum_{i,j=1}^Na^2_{ij}\|f_{ji}(\delta_j(t)-\delta_i(t))\|^2
\leq2\bar{\sigma}^2\Delta_{\Lambda}(t).
$
where $\Delta_{\Lambda}(t)=\overline{\delta}^T(t)(\Lambda \otimes I_n)\overline{\delta}(t)$.
Hence, $$
d \langle M_{\tau_2}\rangle(t)\leq 2\frac{N-1}{N}k^2\bar{\sigma}^2\Delta_{\Lambda}(t-\tau_2)dt.
$$
  Using the It\^{o} formula, we get
  \begin{eqnarray}
   dV(\overline{\delta}_t) &=&2k^2\Big[\int_{t-\tau_1}^t\overline{\delta}(s)ds\Big]^T (\Lambda \otimes I_n)^2\overline{\delta}(t)dt\cr
   &&+ d\langle M_{\tau_2}\rangle(t)
   +  k^2\tau_1\Delta_{\Lambda^2}(t)dt -2k \Delta_{\Lambda}(t)dt \cr
   &&- k^2\int_{t-\tau_1}^t  \Delta_{\Lambda^2}(s)dsdt+dm(t),
  \end{eqnarray}
  where $m(t)=\int_0^t2\Big[\overline{\delta}(s)-k(\Lambda \otimes I_n)\int_{s-\tau_1}^s\overline{\delta}(u)du\Big]^TdM_{\tau_2}(s)$ is a martingale with $m(0)=0$. Using the elementary inequality: $2x^Ty\leq \theta\|x\|^2+\frac{1}{\theta}\|y\|^2$, $\theta>0$, $x,y\in\mathbb{R}^{nN}$, and H\"{o}lder's inequality yields
  $2\Big[\int_{t-\tau_1}^t\delta(s)ds\Big]^T(\Lambda^2 \otimes I_n)\overline{\delta}(t)\leq  \Big(\tau_1 \Delta_{\Lambda^2}(t) +\int_{t-\tau_1}^t\Delta_{\Lambda^2}(s)ds\Big ).$ Note that $\Delta_{\Lambda^2}(t)\leq\lambda_N\Delta_{\Lambda}(t)$.
 Therefore, we have
   \begin{eqnarray}\label{S6dvt}
   dV(\overline{\delta}_t) &\leq&2\frac{N-1}{N}k^2\bar{\sigma}^2\Delta_{\Lambda}(t-\tau_2)dt+ dm(t)\cr
   &&
     -2k(1-\tau_1 k\lambda_N) \Delta_{\Lambda}(t)dt.
  \end{eqnarray}
  Note that for $e^{\gamma t}V(\overline{\delta}_t)$, $\gamma>0$,
 $
e^{\gamma t}\mathbb{E}V(\overline{\delta}_t)=\mathbb{E}V(\overline{\delta}_0)+\gamma\mathbb{E} \int_0^te^{\gamma s}V(\overline{\delta}_s)ds+\mathbb{E}\int_0^te^{\gamma s}dV(\overline{\delta}_s).
 $
  Therefore, we have
  \begin{eqnarray}\label{0519}
   e^{\gamma t}\mathbb{E} V(\overline{\delta}_t) &\leq& \mathbb{E}V(\overline{\delta}_0) +\kappa\mathbb{E}\int_0^te^{\gamma s}\Delta_{\Lambda}(s-\tau_2)ds\cr
   &&-2k(1-\tau_1 k\lambda_N) \mathbb{E}\int_0^t e^{\gamma s}\Delta_{\Lambda}(s) ds\cr
  &&
  +\int_0^t \gamma e^{\gamma s}\mathbb{E}V(\overline{\delta}_s)ds ,
  \end{eqnarray} where $\kappa=2\frac{N-1}{N}k^2\bar{\sigma}^2$.  By the definition of the functional $V(\overline{\delta}_t)$ and the elementary inequality $\|x+y\|^2\leq 2\|x\|^2+2\|y\|^2$, $x,y\in\mathbb{R}^{nN}$,   we have
$
  V(\overline{\delta}_s)  \leq2 \|\overline{\delta}(s)\|^2 +3\lambda^2_Nk^2\tau_1 \int_{s-\tau_1}^s\|\overline{\delta}(u)\|^2 du.
 $
 It is easy to see that
 $\int_0^te^{\gamma s}\Delta_{\Lambda} (s-\tau_2)ds \leq e^{\gamma \tau_2}(\int_{-\tau_2}^0\Delta_{\Lambda}(s)ds+ \int_0^te^{\gamma s}\Delta_{\Lambda}(s)$ $ds).
 $
  Then from \eqref{0519}, we get
   \begin{eqnarray}\label{baobei}
e^{\gamma t}\mathbb{E} V(\overline{\delta}_t) &\leq&  C_1(\gamma) -h_1(\gamma)\int_0^te^{\gamma s} \mathbb{E}\|\overline{\delta}(s)\|^2ds \cr
    &&+\gamma h_2 \int_0^t e^{\gamma s}\int_{s-\tau_1}^s\mathbb{E}\|\overline{\delta}(u)\|^2 du ds,
  \end{eqnarray}
  where $C_1(\gamma)=\kappa e^{\gamma \tau_2} \tau_2 \lambda_{N}\sup_{s\in[-\tau,0]}\mathbb{E}\|\overline{\delta}(s)\|^2+\mathbb{E}V(\overline{\delta}_0)$, $h_1(\gamma)=2 k(1-\frac{N-1}{N}k\bar{\sigma}^2e^{\gamma \tau_2}-\lambda_Nk\tau_1)\lambda_2-2\gamma$, $h_2=3\lambda^2_Nk^2\tau_1.$   Note that
 $
    \int_0^t e^{\gamma s}\int_{s-\tau_1}^s\mathbb{E}\|\overline{\delta}(u)\|^2 du ds
   \leq\tau_1^2e^{\gamma\tau_1} \|\overline{\delta}_0\|_C^2 +\tau_1 e^{\gamma\tau_1}\int_{0}^te^{\gamma u} \mathbb{E}\|\overline{\delta}(u)\|^2du.
  $ 
  This together with \eqref{baobei}  implies
     \begin{eqnarray}\label{geda}
 e^{\gamma t}\mathbb{E} V(\overline{\delta}_t)\leq C_2(\gamma) +h(\gamma)\int_0^te^{\gamma s} \mathbb{E}\|\overline{\delta}(s)\|^2ds,
  \end{eqnarray}
  where $C_2(\gamma)=C_1(\gamma)+h_2\gamma\tau_1^2e^{\gamma\tau_1}\sup_{s\in[-\tau,0]}\mathbb{E}\|\overline{\delta}(s)\|^2$, $h(\gamma)=h_2\gamma\tau_1 e^{\gamma\tau_1}$ $-h_1(\gamma)$.
 It is easy to see from \eqref{s4taudefinition} that  $h(0)=-2 k(1-\frac{N-1}{N}k\bar{\sigma}^2 -\lambda_Nk\tau_1)\lambda_2<0$, and $\lim_{\gamma\rightarrow\infty}h(\gamma)=\infty$.
 Then  there exists a unique positive root, denoted by $\gamma(\tau_2)$, such that $h(\gamma(\tau_2))=0$ and $h(\gamma)<0$ for $\gamma\in(0,\gamma(\tau_2))$. Hence, we get from \eqref{geda} that for $\gamma \in(0,\gamma(\tau_2))$,
 $
 -h(\gamma)\int_0^{\infty}e^{\gamma s} \mathbb{E}\|\overline{\delta}(s)\|^2ds< C_2(\gamma),
 $
 which implies that for certain $C>0$, $
    e^{\gamma   t} \mathbb{E}\|\overline{\delta}(t)\|^2<C.
$
  This together with  the definition of $\overline{\delta}(t)$ produces the mean square weak consensus with a pairwise convergence rate $\gamma$ less than $\gamma(\tau_2)$. It is easy to see that the coefficients in \eqref{s4guiyi}  satisfy a linear growth condition. Then from Theorem 6.2 in \cite[p.175]{Mao2007},
 $
    \limsup_{t\to\infty}\frac{1}{t}\log\|\overline{\delta}(t)\| <-\frac{\gamma}{2}.
$
  By  Lemma \ref{lemconnection}, the agents reach mean square and almost sure strong consensus. Then the remaining is to apply the similar methods used in \citet*{LWZ2014TAC}.

  Now, we prove the necessity of the condition $0<k<\frac{N}{\underline{\sigma}^2(N-1)}$ under $2\tau_2\geq\tau_1$. If $k\geq\frac{N}{\underline{\sigma}^2(N-1)}$, we choose the initial data $x_i(\theta)=x_i(0)$ for $\theta\in[-\tau,0]$. It can be deduced  that
  $$
  d\overline{\delta}(t)=-k(\Lambda\otimes I_n)\overline{\delta}(t-\tau_1)dt+d\widetilde{M}_{\tau_2}(t),
  $$
   where $\widetilde{M}_{\tau_2}(t)=\sum_{i,j=1}^{N}\sigma_{ji}\int_0^t[(\phi^{T} (I_N-J_N)B_{ij}\phi)\otimes K]\overline{\delta}(s-\tau_2) dw_{ji}(s)$, $B_{ji}=[b_{kl}]_{N\times N}$ be an $N\times N$ matrix with $b_{ii}=-a_{ij}, b_{ij}=a_{ij}$ and all other elements being zero, $i,j=1,2,\ldots, N$. Applying the It\^{o} formula, we have $
   d\|\overline{\delta}(t)\|^2 = -2k\overline{\delta}(t)^T(\Lambda \otimes I_n)\overline{\delta}(t-\tau_1)dt+ 2\overline{\delta}(t)^Td\widetilde{M}_{\tau_2}(t)
  +\overline{\delta}(t-\tau_2)^T\Phi_{k}\overline{\delta}(t-\tau_2)dt$, where $\Phi_{k}=k^2\sum_{i,j=1}^N\sigma^2_{ji}(\phi^TB^T_{ij}\phi\phi^TB_{ij}\phi)\otimes I_n$. Using the definition of $\Delta_{\Lambda}(s)$ defined above and noting that $\Lambda \otimes I_n$ is positive definite, then we have $2\overline{\delta}(t)^T(\Lambda \otimes I_n) \overline{\delta}(t-\tau_1)\leq \Delta_{\Lambda}(t)+ \Delta_{\Lambda}(t-\tau_1)$. Hence,
   \begin{eqnarray*}
  \|\overline{\delta}(t)\|^2 &\geq& \|\overline{\delta}(0)\|^2+\int_0^{t}\overline{\delta}(s-\tau_2)^T\Phi_{k}\overline{\delta}(s-\tau_2)ds\cr
   &&  -k \int_{0}^t(\Delta_{\Lambda}(s)+\Delta_{\Lambda}(s-\tau_1))ds+ \breve{M}(t)\cr
    &\geq&-2k \int_{t-\tau_2}^t\Delta_{\Lambda}(s)ds+ \breve{M}(t)+\|\overline{\delta}(0)\|^2\cr
   &&-\int_0^{t-\tau_2}\overline{\delta}(s)^T[2k(\Lambda \otimes I_n)-\Phi_{k}]\overline{\delta}(s)ds \cr
   &&+\overline{\delta}(0)^T[\tau_2\Phi_{k}- \tau_1 k(\Lambda \otimes I_n)] \overline{\delta}(0).
  \end{eqnarray*}
   where $ \breve{M}(t)=2\int_0^t\overline{\delta}(s)^Td\widetilde{M}_{\tau_2}(s).$   By the definition of $B_{ij}$ and $\phi$, we have $\sum_{i,j=1}^N\phi^TB^T_{ij}\phi\phi^TB_{ij}\phi=\frac{2(N-1)}{N}\phi^T\mathcal{L}\phi=\frac{2(N-1)}{N}\Lambda$,
 which together with $K=kI_n$ leads to $2k(\Lambda \otimes I_n)-\Phi_{K}\leq 2 k (1-\frac{k\underline{\sigma}^2(N-1)}{N})(\Lambda\otimes I_{n})$.  Note that $\tau_2\geq\tau_1/2$. We obtain
        \begin{eqnarray*}
\mathbb{E} \|\overline{\delta}(t)\|^2 &  \geq&   \|\overline{\delta}(0)\|^2 -2k \mathbb{E}\int_{t-\tau_2}^t\Delta_{\Lambda}(s)ds\cr
&&\hspace{-10.pt}+  \tau_2  k (\frac{k\underline{\sigma}^2(N-1)}{N}-1)\Delta_{\Lambda}(0)\cr
   &&\hspace{-10.pt}+ 2 k (\frac{k\underline{\sigma}^2(N-1)}{N}-1)\mathbb{E}\int_0^{t-\tau_2}\Delta_{\Lambda}(s)ds.
  \end{eqnarray*}
If $k\geq\frac{N}{\underline{\sigma}^2(N-1)}$, then we have
  $\mathbb{E}\|\overline{\delta}(t)\|^2+2k\lambda_N \int_{-\tau_2}^0\mathbb{E}\|\overline{\delta}(s+t)\|^2ds \geq \|\overline{\delta}(0)\|^2$,
 which implies $\liminf_{t\rightarrow\infty}\mathbb{E}\|\overline{\delta}(t)\|^2>0$ for any given $x(0)$ such that $\overline{\delta}(0)\neq0$. This is in contradiction with the definition of AUMSAC.
  $\Box$

%

%

\begin{remark}\label{rem4.2}
\rm{Note that the sufficient condition \eqref{s4taudefinition} does not involve time-delay $\tau_2$. Hence,  the time-delay $\tau_2$ does not affect the goal of AUMSAC and AUASAC  under the choice of control gain satisfying \eqref{s4taudefinition}. But it may affect  the exponential   convergence
rates $\gamma_{\tau_2}$ and $\gamma_{\tau_2}/2$, and then prolong the time of achieving consensus. In fact,  $\gamma_{\tau_2}$ defined in Theorem \ref{th1} is a decreasing function with respect to $\tau_2$, and satisfies $\lim_{\tau_2\to\infty}\gamma_{\tau_2}=0$. The
simulation examples in Section 5 also confirm the theoretical results. This is also a new interesting  finding in stochastic stability of stochastic delay systems.
}
\end{remark}

\begin{remark} \rm{Theorem \ref{th1} shows that if  the undirected graph $\mathcal{G}$ is connected, then for any fixed $\tau_1,\tau_2\geq0$, the AUMSAC and AUASAC  can be achieved by designing the control gain $K=kI_n$ satisfying \eqref{s4taudefinition}. If the noises disappear, then $\bar{\sigma}^2=0$ and the fixed control gain $K=kI_n$ with $0<k<\frac{1}{\lambda_N\tau_1}$ can ensure deterministic consensus, which is in consistent with Theorem \ref{newthe}.
}
\end{remark}

\section{Simulation examples}\label{S5simu}
We consider  almost sure consensus for a scalar four-agent example under the topology graph $\mathcal{G}=\{\mathcal{V}, \mathcal{E},\mathcal{A}\}$, where $\mathcal{V}=\{1,2,3,4\}$, $\mathcal{E}=\{(1,2), (2,3), (3,4), (4,3), (3,2)\}$ and $\mathcal{A}=[a_{ij}]_{4\times 4}$ with $a_{12}=a_{23}=a_{32}=a_{34}=a_{43}=1$ and other being zero. The initial state  is $x(t)=[-7,4, 3, -8]^T$ for $t\in[-\tau,0]$, $\tau=\tau_1\vee\tau_2$.

\begin{figure}[htbp]
\centering
\includegraphics[width=2.5in]{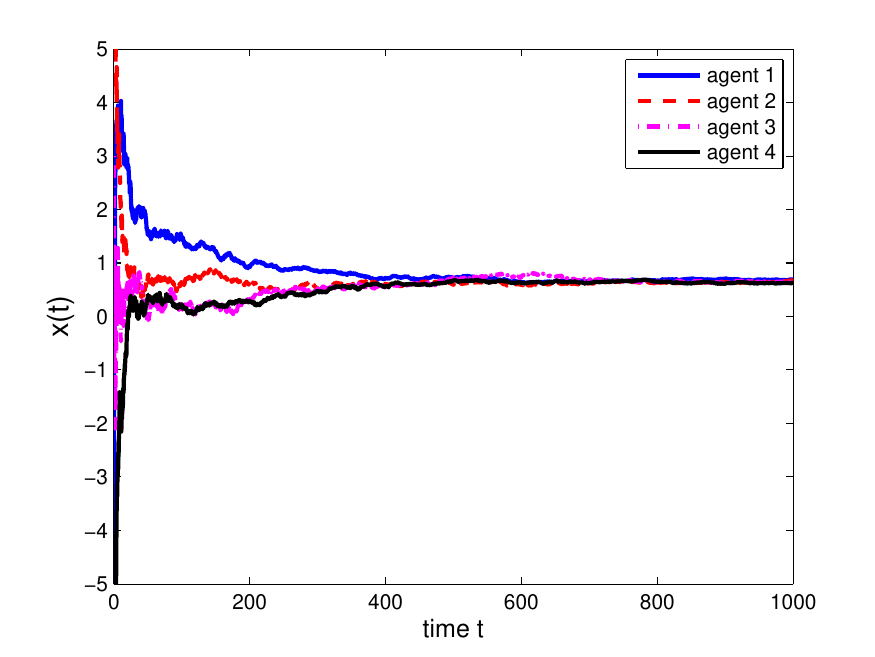}
\caption{States of the four agents with additive noises: $c(t)=(1+t)^{-1}$ and $\tau_1=0.2$.}
\label{figadd1}
\end{figure}

\begin{figure}[htbp]
\centering
\includegraphics[width=2.5in]{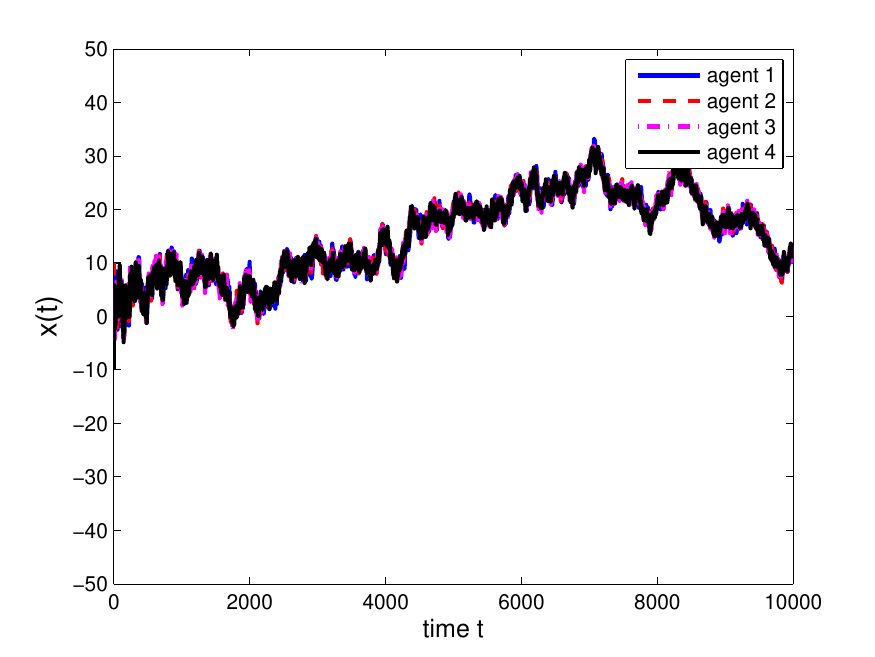}
\caption{States of the four agents with additive noises: $c(t)=(1+t)^{-1/3}$ and $\tau_1=0.2$.}
\label{figadd2}
\end{figure}

\textbf{Additive noise case} It can be seen that the graph  $\mathcal{G}$ contains a spanning tree. Moreover, we can obtain $\lambda_2=\lambda_3=1$ and $\lambda_4= 3$.  Let $\tau_1=0.2$ ($\tau_2=0$)  and $f_{ji}(x)=\sigma_{ji}$ with $\sigma_{ji}=2$, $i,j=1,2,3,4$. We first choose the control gain $c(t)$ as $c(t)=\frac{1}{1+t}$, $t\geq0$, then 
$\tau_1\lambda_4\bar{c}_{t_0}<1$ for any $t_0\geq0$, and
conditions (C1)-(C3) hold. Hence, by Theorem \ref{thadddelay33},   almost sure strong consensus can be achieved, that is, all agents' states will tend to a common value, which is depicted in Fig. \ref{figadd1}.
Then we choose $c(t)=(1+t)^{-1/3}$, $t\geq0$. It is easy to see that condition (C2) is violated, but conditions (C1) and (C5) hold. By Theorem  \ref{thadddelay3}, almost sure weak consensus can be achieved, which is depicted in Fig. \ref{figadd2}. Fig. \ref{figadd2} shows that the agents do not converge to a common value, but they tend to get together in the future, which also shows the necessity of condition (C2) for  almost sure strong consensus. This is  consistent with Theorem \ref{thadddelay33}.

\begin{figure}[htbp]
\centering
\includegraphics[width=2.5in]{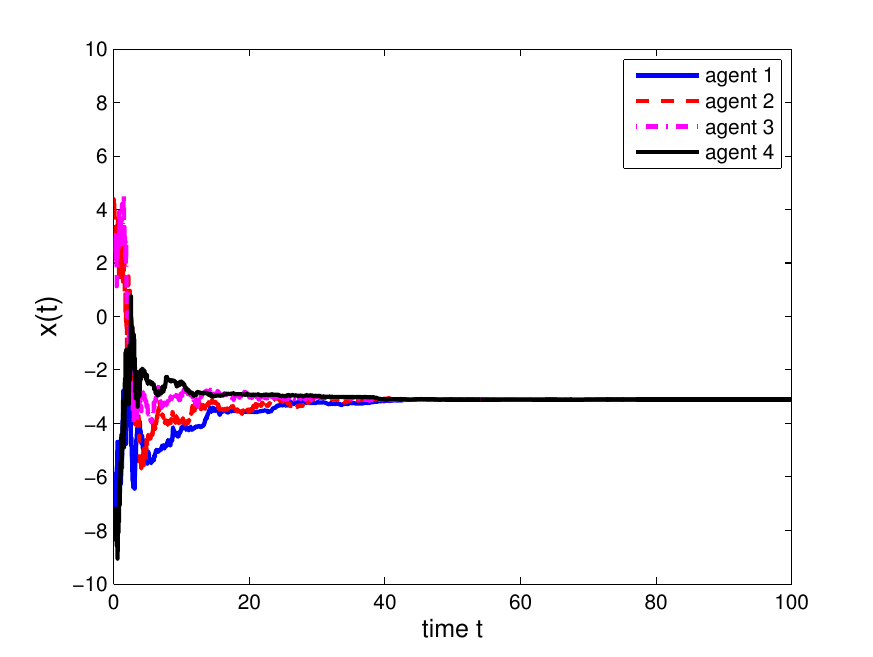}
\caption{States of the four agents  with multiplicative noises: $k=0.12$, $\tau_1=0.2$ and $\tau_2=2$.}
\label{figmulti1}
\end{figure}

\begin{figure}[htbp]
\centering
\includegraphics[width=2.5in]{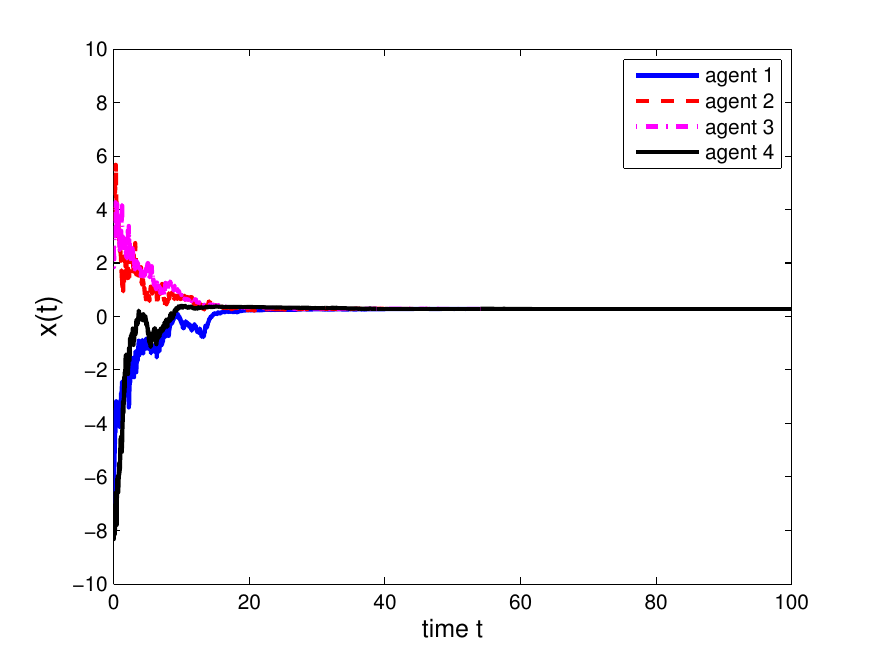}
\caption{States of the four agents  with multiplicative noises: $k=0.12$, $\tau_1=0.2$ and $\tau_2=0$.}
\label{figmulti2}
\end{figure}

\begin{figure}[htbp]
\centering
\includegraphics[width=2.5in]{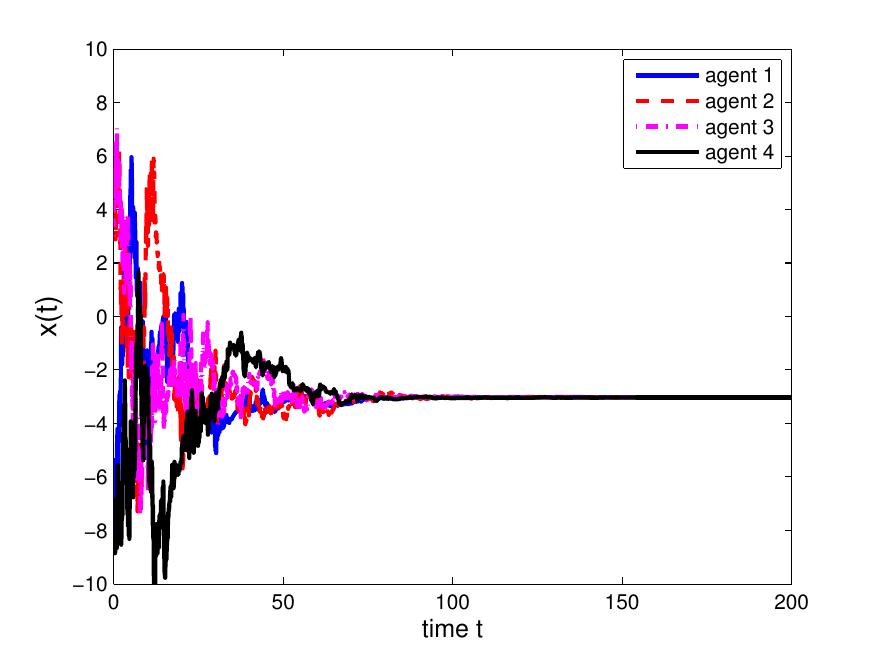}
\caption{States of the four agents   with multiplicative noises: $k=0.12$, $\tau_1=0.2$ and $\tau_2=10$.}
\label{figmulti3}
\end{figure}

\begin{figure}[htbp]
\centering
\includegraphics[width=2.5in]{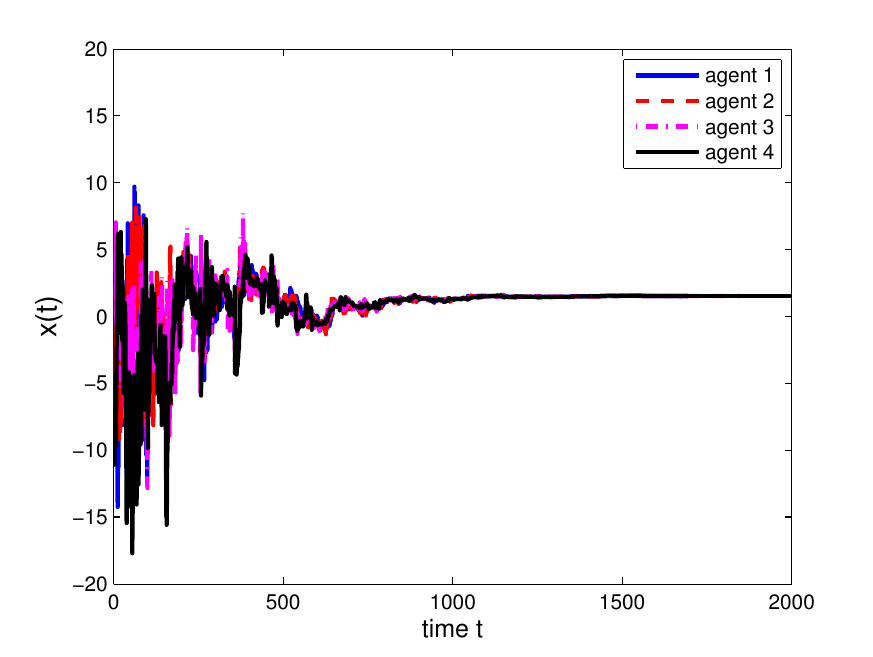}
\caption{States of the four agents  with multiplicative noises: $k=0.12$, $\tau_1=0.2$ and $\tau_2=100$.}
\label{figmulti4}
\end{figure}

\textbf{Multiplicative noise case} Let $a_{21}=1$ and $f_{ji}(x)=\sigma x$, $\sigma=2$, $i,j=1,2,3,4$. Then $\mathcal{G}$ is undirected and $\lambda_2=0.5858$ and $\lambda_4=3.4142$. We first choose the time-delays $\tau_1=0.2$ and $\tau_2=2$, and the control gain $k=0.12<k^*:=\frac{1}{\lambda_4\tau_1+3/4\sigma^2}=0.2715$. Then by Theorem \ref{th1},   almost sure strong consensus can be achieved, which is proved numerically in Fig. \ref{figmulti1}.
Then we aim to examine numerically how the time-delay $\tau_2$ affect the control gain to guarantee almost sure consensus. We choose $\tau_1=0.2, k=0.12$ and $\tau_2=0, 10, 100$, respectively, then we can obtain Figs. \ref{figmulti2}, \ref{figmulti3} and \ref{figmulti4} accordingly. These figures show that time-delay $\tau_2$ does not affect the control gain $k$ to achieve the goal of almost sure consensus, but it may prolong the time of achieving consensus. This confirms  Remark \ref{rem4.2}.

\begin{figure}[htbp]
\centering
\includegraphics[width=2.5in]{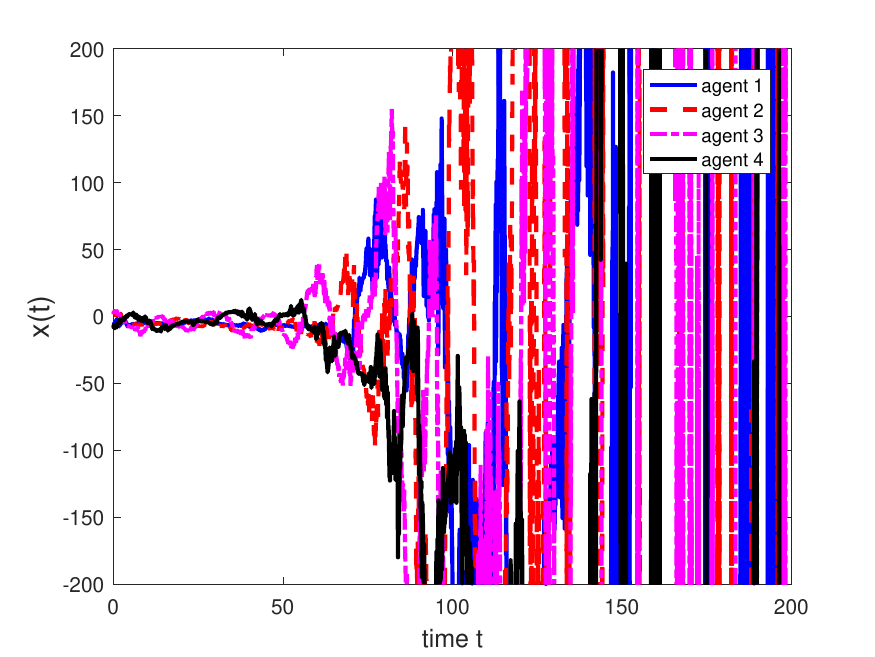}
\caption{States of the four agents   with multiplicative noises: $k=0.12$, $\tau_1=3.5$ and $\tau_2=0$.}
\label{unstab}
\end{figure}

\begin{figure}[htbp]
\centering
\includegraphics[width=2.5in]{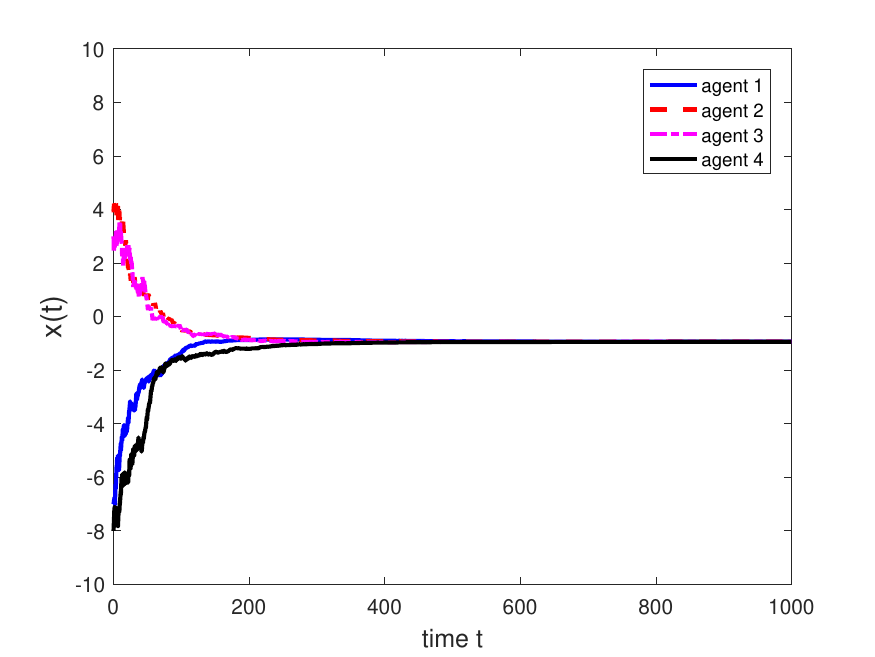}
\caption{States of the four agents   with multiplicative noises: $k=0.013$, $\tau_1=3.5$ and $\tau_2=0$.}
\label{tau1stab}
\end{figure}

We now examine the effect of time-delay $\tau_1$ on the almost sure consensus. Considering $\tau_1=3.5$, we can see that the sufficient condition in  Theorem \ref{th1} is defied for the choice of $k=0.12$ ($>\frac{1}{\lambda_N\tau_1+\frac{N-1}{N} \bar{\sigma}^2}=0.0669$) used above. The simulation in Fig. \ref{unstab} shows that the four agents cannot achieve the almost sure consensus. But if we choose $k_1=0.013$ ($<0.0669$), then the sufficient condition in Theorem  \ref{th1} and the consensus will be achieved, which is revealed in Fig. \ref{tau1stab}.

\section{Conclusion}\label{S6con}
This work addresses  stochastic consensus, including  mean square and  almost sure weak and strong consensus, of high-dimensional multi-agent systems with time-delays and additive or multiplicative measurement noises. The main results are composed of two parts. In the first part, we consider consensus conditions of multi-agent systems with the time-delay and additive noises. Here, the semi-decoupled skill and the differential resolvent
function become the power tools to find the  sufficient conditions for stochastic weak consensus. Then the martingale convergence theorem is applied to obtain  stochastic strong  consensus. The second part takes  time-delays and multiplicative noises into consideration, where the degenerate Lyapunov functional helps us to establish sufficient conditions for  mean square and almost sure strong consensus. 


  Generally speaking, solving almost sure consensus is a more difficult and more challenging work than solving mean square consensus. Moreover, the emergence of time-delay also adds to the difficulty. Although we find the weak conditions for  almost sure consensus under the additive noises, this cannot be extended to the case with multiplicative noises. In Section  \ref{S4multi}, we develop  almost sure consensus based on the conditions of mean square consensus and stochastic stability theorem. However, the similar weak conditions in the delay-free case of \citet{LWZ2014TAC} are difficult to obtain. These issues still deserve further research. In presence of the time-delay and multiplicative measurement noises, this work assumes that the graph is undirected and fixed, and the time-delays in each channel are equal.  In the future works, it would be more interesting and perhaps challenging to consider the general case without these assumptions.  

\section*{Acknowledgements}
This work was supported by the National Natural Science Foundation of China under Grant Nos. 61522310, 61227902, and 61703378, the Shu Guang project of Shanghai Municipal Education Commission and Shanghai Education Development Foundation under grant 17SG26, the Fundamental Research Funds for the Central Universities, China University of Geosciences(Wuhan)(No. CUG170610) and the National Key Basic Research Program of China (973 Program) under Grant No. 2014CB845301.

\end{document}